%% file: RG_in_EFTs.tex
\documentclass[a4paper,twoside,openany,11pt]{memoir} %openany, openright

\title{
    On the Renormalization Group in EFTs: On-Shell Bases, Ambiguities, and Divergences
}
\author{
    Anders Eller Thomsen
}
\def\fullheadfoot{0} % Use 0 for plain foots and 1 for full heads
\input{PreRamble}

\newcommand{\abs}[1]{\left| #1 \right|}
\newcommand{\braces}[1]{\big\lbrace #1 \big\rbrace}

\newcommand{\eminus}{\vcenter{\hbox{\scalebox{0.6}[1]{$ - $}}}}	%Narrow minus signed (for e.g. negative exponents)
\newcommand{\ord}[1]{\mathcal{O}\!\left( #1 \right)}

\newcommand{\hc}{\; + \; \mathrm{H.c.} \;}
\newcommand{\andeq}{\quad \mathrm{and} \quad}

\newcommand{\dd}{\mathop{}\!\mathrm{d}}
\newcommand{\ud}[2]{\phantom{}^{#1}\phantom{}_{#2}}

\newcommand{\sscript}[1]{{\scriptscriptstyle \mathrm{#1}}}

\makeatletter
\newcommand{\vast}{\bBigg@{3}}
\makeatother

\renewcommand{\L}{\mathcal{L}}
\newcommand{\LL}{\mathrm{L}}
\newcommand{\RR}{\mathrm{R}}
\newcommand{\U}{\mathrm{U}}
\newcommand{\SU}{\mathrm{SU}}

\newcommand{\off}{\mathrm{off}}
\newcommand{\on}{\mathrm{on}}
\newcommand{\red}{\mathrm{red}}
\newcommand{\scJ}{\mathcal{J}}
\newcommand{\Ver}{\mathrm{Ver}}

\newcommand{\bef}{$ \beta $-function\xspace}
\newcommand{\befs}{$ \beta $-functions\xspace}
\newcommand{\msbar}{$ \overline{\text{\small MS}} $\xspace}

\begin{document}

% % % % Title % % % % 
\thispagestyle{empty}
\renewcommand*{\thefootnote}{\fnsymbol{footnote}}
\suppressfloats	%Prevents figures and the likes on this page (title page)
% % % % Title % % % % 
\begin{center}
%Title
	{\sffamily \bfseries \mathversion{boldsans} \fontsize{24}{22}\selectfont
%	Running with Infinity
	On the Renormalization Group in EFTs:\\{\fontsize{18}{20}\selectfont On-Shell Bases, Flavor Ambiguities, and Divergences} 
	\\[.005\textheight]
	\textcolor{blue!80!black}{\rule{.25\textwidth}{.7pt}}\\[.03\textheight]}
%Authors
	{\sffamily \mathversion{sans} \Large 
	  	Anders Eller Thomsen\footnote{anders.thomsen@unibe.ch}
	}\\[1.25em]
%Affilitations
	{ \small \sffamily \mathversion{sans} 
	Albert Einstein Center for Fundamental Physics, Institute for Theoretical Physics,\\ University of Bern,  Sidlerstrasse 5, CH-3012 Bern, Switzerland
	}
	\\[.01\textheight]{\itshape \sffamily \today}
	\\[.015\textheight]
	\textcolor{blue!80!black}{\rule{.25\textwidth}{.7pt}}\\[.03\textheight]
\end{center}
\setcounter{footnote}{0}
\renewcommand*{\thefootnote}{\arabic{footnote}}%

\begin{abstract}\vspace{+.01\textheight}	
	Recent results for two-loop renormalization group (RG) functions in effective field theories exhibit unphysical divergences when calculated in an on-shell operator basis. We demonstrate that this can be understood to be a result of omitting non-minimal source terms in the renormalized vacuum functional, which are essential to maintaining renormalizability of and describing the RG flow of Green's functions in an on-shell framework. With the inclusion of the missing source terms, any remaining divergences are ambiguous, generating only unphysical RG flow directed along flavor rotations, and the RG functions are RG-finite. We carefully examine the role of flavor rotations in generating ambiguities in both on- and off-shell RG functions and explore the geometry of a physical coupling space. 
\end{abstract}

\newpage
\section*{Table of Contents}
\toc
\vspace*{.02\textheight}

%\newpage 
% % % % % % % % % % % % % % % % % % % % % % % % % % % % % % % % % % % %
\section{Introduction}
% % % % % % % % % % % % % % % % % % % % % % % % % % % % % % % % % % % %
As the search for new physics beyond the Standard Model (SM) intensifies, effective field theories (EFTs) have emerged as one of the primary tools assisting this exploration in the area of precision physics, just as they have proven indispensable in SM physics. This has led to the development of new methods and tools for phenomenological analysis, geometric interpretations of EFTs, and a new focus on calculations of next-to-leading order Renormalization Group (RG) functions, among other developments. It is in the calculation of EFT RG functions that we find the curiosities that have inspired this work. Half a century since its inception, the RG remains a cornerstone of QFT; nevertheless, fundamental aspects of the mechanism remain largely unexplored. 

From the very introduction of the RG within the framework of dimensional regularization, it was presumed that RG functions are finite (pole-free) in the $ \epsilon= \tfrac{1}{2}(4 -d) $ expansion to facilitate a \emph{finite flow of renormalized Green's functions}. This expected constraint on the RG functions predicts recursive relations between the higher poles of the counterterms and the simple $ \epsilon $-poles~\cite{tHooft:1973mfk}, sometimes referred to as `t Hooft consistency conditions. A counterexample to the consistency conditions was discovered (at three loop order) only after 40 years, in the unbroken phase of the Standard Model, no less. Refs.~\cite{Bednyakov:2014pia,Herren:2017uxn} found that there is an ambiguity in choosing the counterterms---within the same renormalization scheme---and only certain choices lead to finite RG functions. Besides the puzzling emergence of divergent (in the $ \epsilon $-expansion) RG functions, no other issues indicate pathologies with the other counterterm choices. Ref.~\cite{Herren:2021yur} eventually offered a detailed account of the occurrence of divergent RG functions in renormalizable theories. The divergent RG functions still give rise to a finite flow of renormalized Green's functions following the standard Callan--Symanzik equations~\cite{Callan:1970yg,Symanzik:1970rt}: the divergent contributions are proportional to a Ward identity associated with flavor rotations of the theory and cancel in the RG flow. In fact, the ambiguity in choosing counterterms simply incorporates different, unphysical flavor rotations into the RG flow. The condition that the RG functions produce a finite flow of the renormalized Green's functions is called \emph{RG-finiteness}~\cite{Herren:2021yur} and relaxes the condition of separate finiteness for each individual RG function. The ambiguous nature of the RG, including the finite part, was also observed in the framework of the local RG~\cite{Jack:1990eb,Fortin:2012hn,Jack:2013sha}.

Recently, divergent RG functions---\befs and field anomalous dimensions---have been observed in EFTs at the two-loop order~\cite{Jenkins:2023bls,Manohar:2024xbh,Zhang:2025ywe,Naterop:2025cwg} and beyond~\cite{Henriksson:2025vyi}. The divergences are present when the calculations are performed in an \emph{on-shell EFT basis} without redundant \emph{equation-of-motion (EOM) operators}. The field anomalous dimension in the $ \mathrm{O}(N) $ scalar EFT was found to be divergent~\cite{Jenkins:2023bls,Manohar:2024xbh} even in the absence of non-trivial flavor structures in the model. This strongly suggests that the origin of these new divergences in the EFT RG functions is unrelated to the flavor ambiguities that cause divergences in renormalizable theories (and in EFTs, too). It appears to be a new phenomenon altogether. Ref.~\cite{Manohar:2024xbh} argues that the divergence of field anomalous dimensions should be viewed in light of the well-known fact that shifting to an on-shell EFT basis introduces divergences in off-shell Green's functions, meaning that the \emph{vacuum functional}---the generating functional for connected Green's functions---is no longer fully renormalized. By contrast, the on-shell \befs must be finite to account for the finiteness of the physical $ S $-matrix.\footnote{In fact, $ S $-matrix observables depend on the on-shell couplings in a flavor-invariant manner. This opens up the possibility that the on-shell \befs can have a divergent component, as long as it is in the direction of a flavor rotation. We are left with the weaker conclusion that on-shell \befs need only be \emph{RG-finite}. A formulation of a theory in terms of exclusively physical parameters, without remnant flavor ambiguity, is needed for the original argument. However, this would be a very atypical---not to mention impractical---formulation of most theories.}

With this work, we aim to explore the RG in EFTs equipped with an on-shell operator basis. As opposed to off-shell operator bases, on-shell bases do not span a full set of counterterms needed to renormalize a theory. Presumably, one cannot then establish RG functions within this framework that describe the flow of the now divergent Green's functions with a Callan--Symanzik equation. While the \befs of the on-shell couplings (associated with the on-shell basis) are dictated by the invariance of $ S $-matrix observables w.r.t. the renormalization scale, one must ask if it is even meaningful to discuss the field anomalous dimensions, let alone their divergence, in this framework. The on-shell couplings nevertheless carry all physical information of a theory, and a sensible interpretation of the \befs of the on-shell couplings is possible.  

We argue that the divergence of the RG functions is a spurious consequence of neglecting new \emph{non-minimal source terms} (NMSTs) when performing field redefinitions to remove the redundant operators and transition from an off-shell (Green's) basis to an on-shell basis. While these extra source terms can be discarded when computing $ S $-matrix elements~\cite{Georgi:1991ch,Arzt:1993gz,Criado:2018sdb}, they are needed to fully renormalize the vacuum functional in the \emph{on-shell formulation}, the formulation of the theory with an on-shell operator basis. Only with fully renormalized Green's functions are the RG functions, derived from scale invariance of the bare Lagrangian, expected to be RG-finite; that is, they generate a finite flow of the Green's functions. The NMSTs do not demand new counterterms for the on-shell couplings (with minimal counterterm choices) and, thus, neglecting their contribution to the \befs of the on-shell couplings does not cause new divergences. We show with an explicit example how neglecting NMSTs can combine with spurious flavor divergences to produce divergent \befs already at the two-loop order. We wish to emphasize that while the inclusion of NMSTs provides a satisfying resolution to the divergences observed in RG functions, they can safely be ignored in practical calculations aiming to extract the finite piece of the on-shell \befs. 

Our investigation led us to uncover that \befs receive ambiguous contributions in the direction generated by the flavor group whenever a theory is mapped from one coupling space to another. This is particularly prevalent for the on-shell \befs, which are usually determined through a detour to the off-shell coupling space. We propose the introduction of a physical coupling space, obtained by factoring out the flavor ambiguities from the on-shell coupling space, as a non-redundant description of the theory. Only in such a space will the RG functions be unambiguous. Additional work is needed to determine if this physical coupling space has any practical use or if it remains merely a theoretical interpretation of the physical flow of the theory. Regardless, it is important to bear in mind that different \bef calculations may obtain different, but equally correct, results; one must exercise caution in cross-checks.    

The paper opens with a detailed review of field redefinitions in EFTs and how they shape the RG in Section~\ref{sec:prelims}. The role of the flavor group is described in Section~\ref{sec:flavor_group}, which also covers the lesser-known flavor ambiguities of the RG and RG function divergences related to flavor transformations. Section~\ref{sec:NMS} lays out an on-shell framework including NMSTs to obtain a well-behaved, renormalized mapping from off-shell to on-shell EFTs. We show that the on-shell \befs neatly separate from the redundant couplings in this on-shell framework in agreement with previous expectations. To illustrate these points, we examine recent examples from literature~\cite{Manohar:2024xbh,Zhang:2025ywe} in Section~\ref{sec:examples} to demonstrate that (RG-)finite RG functions are recovered in the full on-shell framework, contrary to previous calculations in the truncated on-shell formulation (the usual ``on-shell basis"). Finally, Section~\ref{sec:geometry} examines the geometry of the various coupling spaces and the construction of a physical coupling space. To the extent that it is possible, we have relegated the more advanced mathematics to this section.     
Section~\ref{sec:conclusions} concludes the work.

\section{Preliminaries} \label{sec:prelims}
We begin our discussion by setting up the RG as a flow in the coupling space of an EFT. We discuss the relation between off- and on-shell bases and the standard relations between counterterms and RG functions.

\subsection{Field redefinitions in EFTs}
Let us take, as the subject of our discussion, an EFT constructed from a set of fields collectively denoted $ \eta^I $. Working at a finite order in the EFT expansion, the local operators constructed from fields and their derivatives form a finite vector space $ V_\off $ that comprises all possible off-shell EFTs at that order. We can choose an \emph{off-shell} (Green's) \emph{basis} of gauge-invariant operators $ O_i(\eta) $ by eliminating redundant operators using integration-by-parts relations and algebraic identities. Such a basis fully parameterizes all off-shell Green's functions that are compatible with a local action, and we have $ \mathrm{span} \big(\{O_i\}\big) = V_\off $. The theory space is parametrized by a set of off-shell couplings $ \lambda^i $---including also mass parameters and renormalizable couplings---which fully describe the off-shell theory. The \emph{off-shell Lagrangian} of the theory is then given by
	\begin{equation} \label{eq:Lag_off-shell}
	\L_\off(\eta,\, \lambda) \equiv \L_\mathrm{kin}(\eta) + \lambda^i O_i(\eta),
	\end{equation}
where $ \L_\mathrm{kin} $ are the kinetic terms of the matter fields (which do not contain any couplings).\footnote{To strictly adhere to this parameterization in gauge theories, we may normalize the gauge-kinetic term as $ \eminus \tfrac{1}{4g^2} F_{\mu\nu}^2 \subset \lambda^i O_i$. Then no couplings appear in the covariant derivatives of the matter fields in, e.g., $ \L_\mathrm{kin} = D_\mu \phi^\dagger D^\mu \phi + \ldots $. The discussion here is easily adapted to the more common normalization of the gauge fields.} With a somewhat sloppy notation, we think of $ \lambda^i $ (with indices $ i,j,\ldots  $) as the coordinates of the point $ \lambda = \lambda^i O_i \in V_\off $, with the usual sum over repeated indices. This Lagrangian is perturbatively renormalizable up to the chosen order in the EFT expansion.

It is well-known that the off-shell Lagrangian over-parameterizes the physical theory~\cite{tHooft:1973wag,Georgi:1991ch,Arzt:1993gz,Criado:2018sdb}. Equivalently, we can say that there is redundancy in the Lagrangian description of the theory. Initially it was thought that all operators proportional to the EOMs are irrelevant, but now the redundancy is understood to be invariance of the $ S $-matrix under field redefinitions. Here we consider field redefinitions $ \eta^I \to \xi^I(\eta) $, where $ \xi(\eta) $ is a covariant polynomial of the fields and their derivatives that preserves the structure of~\eqref{eq:Lag_off-shell} (see Note~\ref{note:redefs} for more details).\footnote{A simple example of a field-redefinition in a scalar field theory would be $ \phi \to \phi + \alpha \phi^2 $ for any $ \alpha $. More relevant examples are found in Section~\ref{sec:examples}.} The consequence of the redefinition $ \xi $ is to change the Lagrangian as 
	\begin{equation} \label{eq:off-shell_field_redef}
	\L_\off(\eta,\, \lambda) \longrightarrow \L_\off\big(\xi(\eta),\, \lambda\big) = \L_\off(\eta,\, \lambda|_\xi),
	\end{equation}
which in effect changes couplings $ \lambda^i \to \lambda^i|_\xi $. Dropping higher order terms in the EFT expansion, the change in fields can be completely absorbed as a change in couplings, since $ \L_\off $ fully parameterizes the theory space. 

\begin{note}[note:redefs]{Field redefinitions}
	The path integral is, in principle, invariant under all invertible field redefinitions. In this paper, we consider only redefinitions that preserve the form of the off-shell Lagrangian~\eqref{eq:Lag_off-shell}, i.e., $ \L_\off\big(\xi(\eta),\, \lambda\big) - \L_\mathrm{kin}(\eta) \in V_\off $. Thus, we take $ \xi^I(\eta) $ to be a local function (polynomial in fields and their derivatives). Furthermore, the restriction prohibits constant pieces, which would otherwise introduce tadpole terms, and the linear term should be chosen to preserve the normalization of $ \L_\mathrm{kin} $. All other terms are formally higher-order in the EFT counting---and, thus, perturbative---ensuring that the Jacobian from the change of integration variable vanishes in the path integral~\cite{Criado:2018sdb} in dimensional regularization. The entire field-redefinition can then be absorbed into a change of couplings $ \lambda \to \lambda|_\xi $, except for the introduction of new NMSTs, which will be the subject of Section~\ref{sec:NMS}.
\end{note}

\begin{figure}[t]
	\centering
	\includegraphics{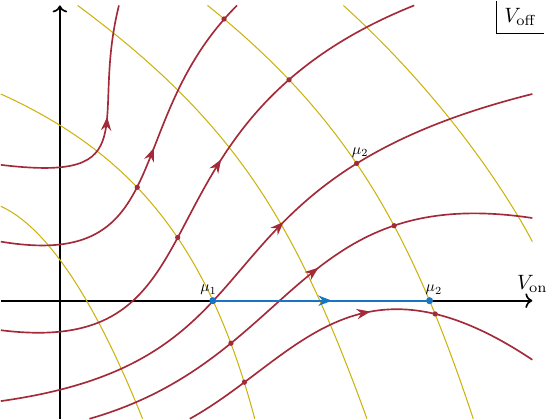}
	\caption{Sketch of the RG flow in the space of off-shell couplings. The horizontal axis indicates the (multi-dimensional) embedding of the on-shell couplings. The yellow lines indicate surfaces that are physically equivalent---the couplings are related by field redefinitions---such that each line correspond to an equivalence class in $ V_\off/\!\sim $. The red lines (with arrows) indicate the flow of the off-shell couplings under the RG. The off-shell flow generated from theory with only on-shell couplings at scale $ \mu_1 $ to the point $ \mu_2 $ will generically bring any theory into the truly off-shell space. Using field redefinitions---following the yellow surface to the intersection with the on-shell subspace---an equivalent on-shell theory is obtained at $ \mu_2 $, from which we infer an ``on-shell'' running of the theory, represented by a blue arrow in $ V_\on $.}
	\label{fig:off-shell_flow}
	\end{figure}

Physics being unchanged under field redefinitions implies an equivalence relation on the space of couplings: we say $ \lambda \sim \lambda' $ if there exists a field redefinition $ \xi(\eta) $ such that $ \lambda'|_\xi = \lambda $. All points in the equivalence class 
	\begin{equation} \label{eq:off-shell_equiv_classes}
	[\lambda] = \braces{\lambda' \in V_\off: \lambda' \sim \lambda}
	\end{equation}
of $ \lambda \in V_\mathrm{off} $ describe the same physics. We stress that the mappings $ \cdot |_\xi $ on the coupling space induced by the field redefinitions are not linear, so the quotient space $ V_\off /\!\sim $ does not preserve the vector space structure of $ V_\off $. The equivalence classes are shown as yellow lines in the sketch of $ V_\off $ shown in Fig.~\ref{fig:off-shell_flow}.\footnote{Relying on equivalence through the leading order EOM  rather than proper field redefinitions would erroneously indicate that $ V_\off /\!\sim $ is a quotient vector space, preserving the linear structure of $ V_\off $ such as in~\cite{Einhorn:2013kja}.} 

The physical redundancy in the off-shell theory space has led to the widespread use of on-shell EFT bases---going back, at least, to the original formulation of the Standard Model Effective Field Theory (SMEFT)~\cite{Buchmuller:1985jz}---which removes the redundancy from the couplings. An on-shell coupling space is a vector subspace of $ V_\off $ of the \emph{smallest possible dimension} that can still capture all possible physics. An on-shell subspace, say $ V_\on \subseteq V_\off $, must satisfy
	\begin{equation} \label{eq:V_on_surjective}
	\forall \lambda \in V_\off \; \exists g \in V_\on: g \in [\lambda],
	\end{equation}
such that $ V_\on $ contains elements from each equivalence class of $ V_\off/\!\sim $. For $ V_\on $ to be an on-shell subspace, any other $ V'\subseteq V_\off $ satisfying~\eqref{eq:V_on_surjective} must have $ \dim V' \geq \dim  V_\on $. This definition is sufficient for now, but we will make it more precise with the introduction of the flavor group in Section~\ref{sec:flavor_group_action}. For the remainder of the article, we will assume that a particular choice of $ V_\on $ has been made. 

We let $ \{ Q_a\} $ be an operator basis for $ V_\on  $ and write the \emph{on-shell} Lagrangian as 
	\begin{equation} \label{eq:Lag_on-shell}
	\L_\on(\eta,\, g) = \L_\mathrm{kin}(\eta) + g^a Q_a(\eta), \qquad g\equiv g^a Q_a \in V_\on.
	\end{equation}
The coordinates $ g^a $ (with indices $ a,b,\ldots  $) are the \emph{on-shell couplings} of the theory.
Conventionally, most EFT calculations are organized so that the on-shell basis operators are a subset of the off-shell basis operators, i.e., $ Q_a = O_a $ for $ 1 \leq a\leq \dim V_\on $, rather than opting for more general linear combinations of the operators. It is usual to refer to the Lagrangian~\eqref{eq:Lag_on-shell} as the EFT in an on-shell basis. From our perspective, it should more properly be referred to as a \emph{truncated on-shell formulation} of the theory, 
so named because it crucially omits the redundant couplings of the NMSTs introduced in~\ref{sec:NMS} and is insufficient to renormalize all Green's functions.\footnote{We use the phrase ``truncated on-shell formulation'' to discriminate it from the regular ``(full) on-shell formulation,'' as introduced in Section~\ref{sec:NMS}. The truncated formulation is what is typically referred to as an EFT in the ``on-shell basis'' in previous literature.}

Given an on-shell basis $ \{ Q_a\} $, there exists, by construction, a coupling-dependent field redefinition $ \eta \to \Xi(\eta,\lambda) $ such that\footnote{In addition to using $ g $ to describe a point in $ V_\on $, with coordinates $ g^a $, we also use $ g: V_\off \to V_\on $ to denote the mapping to the on-shell subspace. We find the double meaning of the symbol to be practical, as $ g $ and $ g(\lambda) $ are both objects in the on-shell subspace. The context should be clear from whether $ g $ has an argument or not.} 
	\begin{equation} \label{eq:implicit_Xi_def}
	\L_\off(\eta,\, \lambda) \longrightarrow \L_\off \big(\Xi(\eta,\lambda),\, \lambda\big) = \L_\on \big(\eta,\, g(\lambda) \big), \qquad g(\lambda) = \lambda|_{\Xi}.
	\end{equation}
In this notation, $ g(\cdot) $ is a non-linear projection from $ V_\off $ to the on-shell subspace $ V_\on $. Any on-shell theory can also be considered as an off-shell theory. There are generally many ways to choose a \emph{right inverse} (a \emph{section}) $ \lambda(g) $ such that $ g(\lambda(g)) = g $. A natural, even ubiquitous, choice in EFT calculations is to use the trivial embedding $ V_\on \subseteq V_\off $ and let $ \lambda(g) = g $. In principle, this is just a matter of convenience rather than necessity. We will see in Section~\ref{sec:projection_nonuniqueness} that $ \Xi $ and, thus, $ g(\lambda) $ are unique only up to a coupling-dependent flavor rotation.

\subsection{RG flow in theory space} \label{sec:beta_functions}
The off-shell Lagrangian~\eqref{eq:Lag_off-shell} allows for the calculation of physical observables ($ S $-matrix elements) $ \mathcal{O}(\lambda, \mu) $ in terms of the theory parameters and the renormalization scale $ \mu $. Since observables are independent of field redefinitions in the theory, they can equivalently be written in terms of the less redundant on-shell couplings: 
	\begin{equation}
	\mathcal{O}(\lambda, \mu) = \mathcal{O}\big(g(\lambda), \mu\big).
	\end{equation}
This observation provides valuable insight into the nature of the RG. Observables, viewed as a function of the on-shell couplings, are renormalization scale independent, so they satisfy the RG equation 
	\begin{equation} \label{eq:obs_RG_invariance}
	\left(\mu \dfrac{\partial}{\partial \mu} + \beta_\on^a \partial_a  \right)\!  \mathcal{O}(g,\, \mu) = 0, \qquad \beta_\on^a \equiv \dfrac{\dd g^a}{\dd t}, \qquad \partial_a \equiv \dfrac{\partial }{\partial g^a},
	\end{equation}
where $ t= \ln \mu $ is the RG time. Since the observables are functions only of the on-shell couplings, so too are the coupling \befs $ \beta_\on^a = \beta_\on^a(g) $. The physical theory, and its flow, is fully determined by the on-shell couplings~\cite{Einhorn:2001kj}.

The \befs in the off-shell theory can be computed with standard methods~\cite{tHooft:1973mfk}: parametrizing the bare (RG-invariant) couplings in terms of the renormalized ones, the \befs are related to the coupling counterterms (more details to follow in Section~\ref{sec:cts_and_rgs_off-shell}). We write 
	\begin{equation}
	\beta_\off^i(\lambda)\equiv \dfrac{\dd \lambda^i}{\dd t}, 
	\end{equation}
for the off-shell \befs.
Both renormalized Green's functions and observables $ \mathcal{O}(\lambda) $ are RG invariant in the off-shell theory---satisfying, e.g., an off-shell version of~\eqref{eq:obs_RG_invariance}---and the \befs give rise to a flow of the couplings in the space $ V_\off $ as sketched in Fig.~\ref{fig:off-shell_flow} (see also the discussion in Note~\ref{note:linear_order}). For physics to be consistently described by all points in the same equivalence class, it follows that all points in an equivalence class will flow to points residing in another shared equivalence class after the same change in renormalization scale. Were this not to be the case, two different descriptions of the same physics (two points in the same equivalence class) would lead to different physics at another renormalization scale, indicating some sickness in our EFT framework.\footnote{This is assuming that no two distinct equivalence classes in $ V_\off/\!\sim $ produce the same physics at a fixed renormalization scale. Otherwise one would have to identify larger equivalence classes than those obtained by field redefinitions.} This is illustrated by the red points at the $ \mu_1 $ and $ \mu_2 $ equivalence surfaces in the figure. While one might recover on-shell \befs with a similar calculation, bringing the theory to the (truncated) on-shell basis generally compromises the renormalized vacuum functional. The resulting loss of renormalizability indicates that the RG functions will fail to produce a flow for the Green's functions of the theory. We will discuss how to remedy this in Section~\ref{sec:NMS}.  

	\begin{note}[note:linear_order]{Non-decoupling of EOM-type operators}
	One might wonder if it is possible to ``straighten out'' the equivalence classes and RG flows sketched in Figure~\ref{fig:off-shell_flow} by more carefully selecting the operator spaces. We think of the vertical space $ V_\perp $ satisfying $ V_\off = V_\on \oplus V_\perp $ as being spanned by redundant operators (removable by field redefinitions). This is a largely arbitrary choice and we do not have a sense of perpendicularity in $ V_\off $, as there is no inner product on the vector space. 
	It is well-known that EOM operators, of the form 
		\begin{equation*}
		E_p(\eta) = O_{pI}(\eta) \dfrac{\delta S_{\mathrm{ren}}[\eta]}{\delta \eta^I(x)},
		\end{equation*}
	for polynomials $ O_{pI}(\eta) $ of couplings and derivatives, vanish at linear order when inserted into physical matrix elements. Here $ S_{\mathrm{ren}} $ denotes the renormalizable part of the action, meaning that $ E_p $ actually depends on the renormalizable couplings. (This is incompatible with our definitions, but we ignore this issue for now.) Thus, choosing a basis of $ E_p $ to span $ V_\perp $ would seem to render the physical equivalence classes parallel to $ V_\perp $. Furthermore, the EOM-type operators do not mix into the physical on-shell operators under the RG at leading EFT order~\cite{Kluberg-Stern:1975ebk,Deans:1978wn,Collins:1984xc}. However, this whole construction only works at the first non-trivial EFT order beyond the renormalizable Lagrangian: the non-linearity associated with the field redefinitions render this construction infeasible beyond that.
	
	It is possible to construct a simple example to show that even a theory consisting purely of EOM operators is not equivalent to the free theory beyond linear order in the EFT expansion. Consider the toy-model 
		\begin{equation*}
		\mathcal{L} = \tfrac{1}{2} (\partial_\mu \phi_1)^2 + \tfrac{1}{2} (\partial_\mu \phi_2)^2 + c\, \phi_1 \phi_2  \partial^2 \phi_1
		\end{equation*}
	with two real scalar fields. The only coupling is for an EOM operator, which is obviously removable with a field redefinition. Accordingly, we may apply the field redefinition $ \phi_1 \to \phi_1 + c\, \phi_1 \phi_2 + c^2 \phi_1 \phi_2^2 $ to remove said operator (and another EOM operator, appearing at the next order):
		\begin{equation*}
		\mathcal{L} = \tfrac{1}{2} (\partial_\mu \phi_1)^2 + \tfrac{1}{2} (\partial_\mu \phi_2)^2 + \dfrac{c^2}{2} \phi_1 \phi_2  \partial^2 (\phi_1 \phi_2) + \mathcal{O}\big(c^3\big).
		\end{equation*}
	The new operator at order $ c^2 $ is not fully removable, indicating that $ V_\perp \notin V_\off/\!\sim $ even when we let it consist of EOM operators.
	\end{note}

One alternative way to RG-evolve an on-shell theory is to embed it in $ V_\off $ first, then evolve the theory off-shell with the off-shell RG functions, before projecting back to $ V_\on $ (see, e.g.,~\cite{Manohar:2024xbh}). Since the off-shell RG is well-defined, this describes a satisfactory prescription for the on-shell coupling flow. We have sketched this procedure in Fig.~\ref{fig:off-shell_flow}.
Starting from the blue on-shell point at scale $ \mu_1 $, the theory is evolved following the off-shell running (red arrow) to the scale $ \mu_2 $. Then one follows the yellow surface of physically equivalent theories back to the intersection with the on-shell subspace.\footnote{For simplicity, the intersection of the equivalence class with $ V_\on $ is shown as a point in Fig.~\ref{fig:off-shell_flow}. Typically, it is actually a multi-dimensional surface generated by the action of the flavor group (cf. Fig.~\ref{fig:flow_projection}).} This procedure describes an equivalent flow on the on-shell subspace shown in blue. Linearizing this circuitous route, considering an infinitesimal change in RG time, gives 
	\begin{equation}
	\begin{aligned}
	g^a + \! \dd t\, \beta_\on^a(g)  &= g^a \big(\lambda^i(g) + \! \dd t\, \beta_\off^i(\lambda(g)) \big) \\
	&= g^a + \! \dd t\, \beta_\off^i(\lambda(g)) \, \partial_i g^a(\lambda(g)),
	\end{aligned}
	\qquad\qquad  \partial_i \equiv \dfrac{\partial}{\partial \lambda^i}.  
	\end{equation}
Equating the linear terms yields
	\begin{equation} \label{eq:mapping_on/off-shell_bef}
	\beta_\on^a(g) = \beta_\off^i(\lambda) \dfrac{\partial g^a(\lambda)}{\partial \lambda^i}  \bigg|_{\lambda = \lambda(g)}
	\end{equation}
for the on-shell \befs as a function of on-shell couplings. While the on-shell \befs are sometimes thought of as physical, we caution the reader that there is remnant redundancy in this formulation (cf. Section~\ref{sec:flavor_group}). As long as the off-shell \befs are finite, then so are the on-shell ones. While this procedure works well, it is not very practical for calculating the on-shell \befs: it requires access to an off-shell basis (which can be difficult to obtain) and one needs to calculate counterterms for many more couplings than are in the on-shell basis, complicating calculations unnecessarily.

\subsection{Counterterms and RG functions} \label{sec:cts_and_rgs_off-shell}
RG functions are ultimately derived from counterterms in dimensional regularization. The off-shell Lagrangian~\eqref{eq:Lag_off-shell} is renormalized by substituting the renormalized couplings for bare ones:\footnote{One should also add gauge-fixing and ghost terms to the Lagrangian (if the theory is gauged), to make it ready for perturbative calculations. This might also include the introduction of additional gauge-variant counterterms unless a background field gauge is utilized. However, the \befs of $ \lambda^i $ are independent on the chosen gauge~\cite{Caswell:1974cj} and we will not further discuss the subtleties of gauge fixing.}
	\begin{equation} \label{eq:Lag_off-shell_w/_source}
	\L_\off(\eta,\, \lambda,\, \scJ) \equiv \L_\mathrm{kin}(\eta) + \lambda^i_{\mathrm{b}} O_i(\eta) +\scJ_{\mathrm{b},I} \eta^I.
	\end{equation}
At this point we also explicitly introduce the bare field sources $ \scJ_{\mathrm{b},I} $ with which we source the connected Green's functions in the vacuum functional. The bare couplings are split into renormalized couplings $ \lambda^i $ and associated counterterms $ \delta \lambda^i(\lambda) $ to render the vacuum functional a finite function of the renormalized couplings. 
We work in the MS/\msbar scheme\footnote{The \msbar scheme is obtained from MS by substituting the renormalization scale for $ \bar{\mu} $, letting $ \mu^2 = (4\pi)^{\eminus 1}\bar{\mu}^2 e^{\gamma_\mathrm{E}} $. This distinction is irrelevant for the discussion in this paper.} for the remainder of the paper to keep the discussion minimal; we have no reason to suspect that our general conclusions change for other renormalization schemes. In the minimal subtraction (MS) scheme  
	\begin{equation} \label{eq:bare_cts}
	\lambda_{\mathrm{b}}^i = \mu^{k_i \epsilon}\big(\lambda^i + \delta \lambda^i \big), \qquad \delta \lambda^i 
	= \sum_{\ell = 1}^{\infty} \dfrac{\delta^{(\ell)} \! \lambda^i}{(16\pi^2)^\ell }
	=  \sum_{n=1}^{\infty} \dfrac{\delta_n \! \lambda^i}{ \epsilon^n}
	= \sum_{\ell = 1}^{\infty} \sum_{n=1}^{\ell} \dfrac{\delta^{(\ell)}_n \! \lambda^i}{(16\pi^2)^\ell \epsilon^n}.
	\end{equation}
The renormalization scale $ \mu $ is introduced to ensure that the renormalized couplings have the dimensions of the couplings in the unregulated (four-dimensional) theory.\footnote{The dimensionality $ k_i $ is, for instance, $ k_i=1 $ for Yukawa couplings and $ k_i = 2 $ for quartic couplings. The index of $ k_i $ does not count as a repeated index for the purposes of Einstein summation.} Eq.~\eqref{eq:bare_cts} also introduces notation for the counterterms as loop, pole, or mixed expansions, which will be needed in due course. 

Renormalization of the couplings is sufficient to renormalize all subdivergences and vacuum graphs; however, a source renormalization is needed to renormalize $ n $-point Green's functions (cf. Note~\ref{note:source_renormalization}). A multiplicative renormalization factor $ Z\ud{I}{J}(\lambda) $ is sufficient to this end, and we have
	\begin{equation} \label{eq:Z_parameterization}
	\scJ_{\mathrm{b},I} =  \scJ_J (Z^{\eminus 1})\ud{J}{I}, \qquad Z  
	= \mathds{1} + \sum_{\ell = 1}^{\infty}  \dfrac{Z^{(\ell)} }{(16\pi^2)^\ell }
	= \mathds{1} + \sum_{n=1}^{\infty} \dfrac{Z_n }{ \epsilon^n}
	= \mathds{1} + \sum_{\ell = 1}^{\infty} \sum_{n=1}^{\ell} \dfrac{Z^{(\ell)}_n }{(16\pi^2)^\ell \epsilon^n}.
	\end{equation}
The source and coupling counterterms span all possible counterterms of the off-shell action up to the given order in the EFT expansion. Any additional structures are removable via field redefinitions to bring the Lagrangian back to the form~\eqref{eq:Lag_off-shell_w/_source} and are thus not needed to renormalize the off-shell Green's functions.

	\begin{note}[note:source_renormalization]{Source renormalization contra wave-function renormalization}
	To maintain the canonical normalization of the kinetic term, we adopt the convention of omitting the field wave-function normalization in favor of renormalizing the sources, effectively introducing the notion of a bare source $ \scJ_\mathrm{b} $.
	While practical multi-loop computations subtracting subdivergences with variations of the $ \boldsymbol{R} $-operation~\cite{Chetyrkin:2017ppe} may find it convenient to use counterterms for the kinetic terms rather than the field sources, the renormalization factors can always be shifted back after the fact with a field redefinition (not of the type in Note~\ref{note:redefs}, as it explicitly changes the normalization of $ \L_\mathrm{kin} $). Our approach has the benefit that it does not obscure the anti-Hermitian part of the normalization factor $ Z\ud{I}{J} $. Additionally, it does not rely on a dubious introduction of a ``bare field,'' even as the field is an integration variable of the path integral. 
	\end{note}

The RG equations are derived on the basis that bare quantities in the off-shell Lagrangian remain invariant under variations of the renormalization scale. This relies on there being a bare coupling or source for every counterterm operator needed to renormalize the off-shell vacuum functional; it is, for instance, well known that if one omits symmetry-allowed couplings in renormalizable Lagrangians, they will be generated during the RG flow, invalidating the original omission. 
Only with the inclusion of all bare couplings will the derived RG functions describe the flow of the renormalized vacuum functional through the Callan--Symanzik equation [cf.~\eqref{eq:CS}]. 
Failure to include all couplings will also result in divergent RG functions---equivalently, violations of `t Hooft consistency conditions---even at parameter points where some renormalized couplings vanish.

As an ansatz for the coupling \befs, we use a series in $ 1/\epsilon $ (combined with a loop expansion for later use):
	\begin{equation} \label{eq:beta_parameterization}
	\beta_\off^i(\lambda) \equiv \dfrac{\dd \lambda^i }{\dd t} = \sum_{\ell = 0}^{\infty} \dfrac{\beta_\off^{(\ell),i}}{(16\pi^2)^\ell} =\sum_{n= \eminus 1}^{\infty} \dfrac{\beta^i_{\off,n}}{\epsilon^n}= \sum_{\ell = 0}^{\infty} \sum_{n= \eminus 1}^{\ell-1} \dfrac{\beta^{(\ell),i}_{\off,n}}{(16\pi^2)^\ell \epsilon^n}
	\end{equation}
Plugging the series into the equation obtained from RG invariance of the bare couplings~\eqref{eq:bare_cts} ($ 0= \dd \lambda_{\mathrm{b}}^i/\! \dd t$), yields a recursive system of equations (see, e.g.,~\cite{Herren:2021yur}) from which one obtains $ \beta_{\off,\eminus 1}^i = -k_i \lambda^i $ and 
	\begin{equation} \label{eq:beta_formula_off-shell}
	\beta_{\off,n}^i = (\zeta - k_i) \delta_{n+1} \lambda^i  - \sum_{m=0}^{n-1} \beta_{\off,m}^j \partial_j \delta_{n-m} \lambda^i, \qquad n\geq 0.
	\end{equation}
From here we may determine each $ \beta_{\off,n}^i $ from the $ \delta_{n+1} \lambda^i $ counterterm and the lower-pole \befs and counterterms. $ \zeta = k_i \lambda^i \partial_i $ is loop-counting operator\footnote{This follows from the relations between vertices and the loop order of 1PI Feynman graphs.} such that $ (\zeta - k_i) \delta^{(\ell)} \! \lambda^i = 2 \ell \,\delta^{(\ell)}\! \lambda^i $. Typically, $ \beta_{\off,0}^i $ is referred to simply as the \bef: it describes the finite flow of the renormalized couplings. The \emph{invalid assertion} that all poles vanish implies (`t Hooft) consistency relations between higher poles of the counterterms [taking the l.h.s. of~\eqref{eq:beta_formula_off-shell} to zero for $ n> 0 $]. As we will see in Section~\ref{sec:RG_divergence}, this condition is not necessarily satisfied, even in completely healthy theories. 

Another common RG function is the field anomalous dimension $ \gamma $.  It is defined by\footnote{The field anomalous dimension dictates the running of the renormalized sources: \[\dfrac{\dd \scJ_I}{\dd t} = \scJ_J \gamma\ud{J}{I}.\]}
	\begin{equation} \label{eq:gamma_def}
	\gamma^\off(\lambda) \equiv Z^{\eminus 1} \dfrac{\dd }{\dd t } Z
	= \sum_{\ell = 0}^{\infty} \dfrac{\gamma^{\off (\ell)}}{(16\pi^2)^\ell }
	= \sum_{n= 0}^{\infty} \dfrac{\gamma^\off_{n}}{ \epsilon^n} = \sum_{\ell = 0}^{\infty} \sum_{n= 0}^{\ell-1} \dfrac{\gamma^{\off(\ell)}_{n}}{(16\pi^2)^\ell \epsilon^n}
	\end{equation}
again allowing for poles in the $ \epsilon $-expansion. The renormalization factor $ Z $ depends on the renormalization scale only indirectly, through the renormalized couplings. Substituting the expansions of the wave-function renormalization factor~\eqref{eq:Z_parameterization} and the \befs~\eqref{eq:beta_parameterization} produces a recursive formula for the poles of $ \gamma $: 
	\begin{equation}
	\gamma^\off_{n} = \zeta Z_{n+1} + \sum_{m=0}^{n-1} \left[\beta_{\off,m}^j \partial_j Z_{n-m} - Z_{n-m} \gamma^\off_{m} \right], \qquad n\geq 0,
	\end{equation}
where the loop-counting operator yields $ \zeta Z^{(\ell)} = 2 \ell Z^{(\ell)} $. As with \befs, it is common to refer to the finite part $ \gamma^\off_{0} $ as the field anomalous dimension, since it produces a finite flow of the Green's functions. The false assumption that all poles vanish would then give rise to consistency conditions for the higher pole renormalization factors $ Z_{n\geq 2} $. 

The derivation of the relations between counterterms and RG functions has focused on the off-shell formulation of the theory up to this point. It is not evident that a similar derivation of the RG functions will work in the truncated on-shell formulation of the theory. The Lagrangian of the truncated on-shell formulation is the naive bare version of the Lagrangian~\eqref{eq:Lag_on-shell},
	\begin{equation} \label{eq:on-shell_bare_lag}
	\L_\on(\eta,\, g,\, \scJ)  \stackrel{?}{=} \L_\mathrm{kin}(\eta) + g_{\mathrm{b}}^a Q_a(\eta) + \scJ_{\mathrm{b},I} \eta^I.
	\end{equation}
While one might assert that this bare Lagrangian is RG invariant, it is not clear what this assertion means for other aspects of the theory; after all, this Lagrangian is known to fail to renormalize the vacuum functional. There would be no reason to expect the resulting RG functions to govern the flow of renormalized Green's functions.\footnote{The \befs obtained this way, nevertheless, govern the flow of physical observables, which are renormalized by~\eqref{eq:on-shell_bare_lag}~\cite{Criado:2018sdb}. Our analysis supports the notion that the on-shell \befs obtained in the truncated on-shell scheme are sufficiently well-behaved for this to be the case.} 
One could attempt to determine the counterterms of the truncated on-shell formulation directly by demanding that observables are finite; however, a more reasonable and practical approach would be to utilize the embedding of $ V_\on $ in $ V_\off $.
While the embedding $ \lambda(g) $ and projection maps $ g(\lambda) $ are finite for finite (renormalized) couplings, nothing prohibits their use for bare couplings with the implied $ \epsilon $-pole expansion. The embedding of the renormalized couplings $ \lambda(g) $ is valid for setting the off-shell renormalized couplings. Renormalization of the off-shell theory dictates that $ \lambda_{\mathrm{b}}(\lambda(g)) $ renormalizes this point in parameter space. A projection of these bare off-shell couplings back to $ V_\on $ should then recover suitable bare couplings of the truncated on-shell formulation: 
	\begin{equation}
	g_{\mathrm{b}}^a(g) = g^a(\lambda(g_\mathrm{b})) = g^a(\lambda_{\mathrm{b}}(\lambda(g))).
	\end{equation}
A pole expansion of the map in terms of renormalized couplings and counterterms yields 
	\begin{equation} \label{eq:on-shell_counterterms_simple_pole}
	g^a + \delta g^a = g^a(\lambda + \delta \lambda(\lambda(g) ) ) =  g^a + \dfrac{\partial g^a(\lambda)}{\partial \lambda^i} \dfrac{\delta_1 \lambda^{i}}{\epsilon} \bigg|_{\lambda = \lambda(g)} + \mathcal{O}\big(\epsilon^{\eminus 2}\big),
	\end{equation}
which allows for easy extraction of the simple poles of the on-shell counterterms. 

Assuming that the \bef formula~\eqref{eq:beta_formula_off-shell} is straightforwardly applicable in the truncated on-shell formulation (substituting off-shell for on-shell couplings), the finite part of the \bef, $ \beta_{\on,0}^a(g) $, is fully determined by the simple pole counterterm. In that event, the on-shell counterterms~\eqref{eq:on-shell_counterterms_simple_pole} reproduce the finite part of the \befs~\eqref{eq:mapping_on/off-shell_bef} for the on-shell couplings, obtained by embedding the on-shell couplings in the off-shell space and RG evolving there. The construction of the on-shell counterterms and finite \befs provides a computationally cheap way of determining valid on-shell \befs. In this approach, one only has to compute the contribution to off-shell counterterms from the on-shell couplings rather than from all off-shell couplings. This approach to the calculation of on-shell \befs has been widely used in the literature and gives a consistent value for the finite flow; however, it is an incomplete procedure for the full set of RG functions, which is the origin of the inconsistencies in the $ \epsilon $-pole part of the RG functions observed in~\cite{Jenkins:2023bls,Manohar:2024xbh,Zhang:2025ywe,Naterop:2025cwg}. In section~\ref{sec:NMS}, we will show how to adapt the formulas for the RG functions properly to the \emph{(full) on-shell formulation} of the theory by including NMSTs. This inclusion allows for consistently computing the poles of the RG functions. As we will now see, some $ \epsilon $-poles of RG functions are fully consistent and not attributable to a failure of any theory formulation.

\section{The Flavor Group} \label{sec:flavor_group}
So far we have ignored an important piece of structure in the coupling spaces, both off- and on-shell. We now look to examine the consequences of unitary field transformations for the RG and the origin of potential flavor divergences of the RG functions.

\subsection{The action of the flavor group} \label{sec:flavor_group_action}
The key structure organizing the coupling space is the \emph{flavor group} $ G_F $, defined as the largest, unitary symmetry group of $ \L_\mathrm{kin}(\eta) $. It is so named because it takes the form of a flavor rotation---a unitary transformation that rotates fields sharing the same gauge and Lorentz group quantum numbers. The fields transform in some linear representation of $ G_F $, and we write the group action as $ a \cdot \eta $ for any $ a \in G_F $. By definition, then, $ \L_\mathrm{kin}(a\cdot \eta) = \L_\mathrm{kin}(\eta) $.

Although we have been somewhat vague about the definition of the basis operator $ O_i(\eta) $ up to this point, our implicit assumption is that they are (and have been) flavor covariant monomials in fields and their derivatives, meaning that under transformations $ a\in G_F $, we have
	\begin{equation}
	O_i(a\cdot \eta) = (a\cdot O)_i(\eta). 
	\end{equation}
Since the fields transform under a linear representation of $ G_F $, it follows that the group action on $ O_i $ is linear too. This endows the off-shell couplings with a transformation under a representation of $ G_F $:
	\begin{equation}
	a\cdot \lambda \equiv (a \cdot \lambda)^i O_i \in V_\off, \qquad a \in G_F,\; \lambda \in V_\off, 
	\end{equation}
such that $ \lambda^i O_i( a^{\eminus 1}\cdot \eta) = (a \cdot \lambda)^i \cdot O_i (\eta) $.\footnote{$ O_i $ and $ \lambda^i $ transform under conjugate representations.} 
The Lagrangian $ \L_\off $ is invariant under a simultaneous transformation of fields, couplings, and sources. The on-shell subspace $ V_\on\subseteq V_\off $ transforms under a sub-representation of $ G_F $, meaning that the action closes:
	\begin{equation}
	a \cdot g \in V_\on, \qquad  \forall a \in G_F,\, g \in V_\on.
	\end{equation} 

The action of the flavor group on $ V_\off $ can also be understood to be a change resulting from a unitary field redefinition $ \eta^I \to \xi^I(\eta) = (a^{\eminus 1} \cdot \eta)^I $, which is a subset of the field redefinitions introduced in Note~\ref{note:redefs}. Indeed we have
	\begin{equation}
	\L_\off\big(\xi(\eta), \lambda \big) = \L_\mathrm{kin}(\eta) + \lambda^i O_i(a^{\eminus 1}\cdot \eta) =\L_\mathrm{kin}(\eta) + (a\cdot \lambda)^i O_i(\eta) = \L_\off(\eta, a\cdot \lambda).
	\end{equation}
This type of field redefinitions does not allow for writing down fewer flavor-covariant operators in the Lagrangian nor is it needed for reducing an EFT to an on-shell basis.

Returning to our definition of the on-shell space $ V_\on $, the definition allows for the possibility that multiple parameter points describe the same physics. Clearly, a unitary transformation of fields does not move outside the $ V_\on $ subspace, although it generally acts non-trivially on the couplings. With the introduction of the flavor group, we assert that physically equivalent on-shell points [in the same equivalence class~\eqref{eq:off-shell_equiv_classes}] are related by flavor transformations:\footnote{Given the non-linearity of the equivalence relation, the existence of such a $ V_\on $ is perhaps not obvious. It is, however, well-established that field redefinitions can systematically remove all kinetic-EOM operators (operators proportional to $ \delta S_\mathrm{kin}[\eta]/\delta \eta $) from a Lagrangian~\eqref{eq:Lag_off-shell} using the perturbative ordering of the EFT expansion. Once this is done, any additional non-unitary redefinition reintroduces EOM operators. Any equivalence class in $ V_\off /\!\sim $ contains one $ G_F $-orbit free of kinetic-EOM operators, and so we can take $ V_\mathrm{on} $ to be the span of all basis operators free of kinetic-EOM operators~\cite{Criado:2018sdb,Fuentes-Martin:2022jrf}. This construction presupposes that $ \{ O_i\} $ has been chosen to maximize the set of kinetic-EOM operators. Other basis choices, such as for instance in the strongly-interacting light Higgs effective Lagrangian~\cite{Giudice:2007fh}, keep kinetic-EOM operators in favor of removing other operators.} 
	\begin{equation} \label{eq:V_on_injective}
	\forall g, g'\in V_\on\!:\, [g'] = [g] \Longleftrightarrow  g' \in G_F\cdot g,
	\end{equation}
where $ G_F\cdot g = \big\{a\cdot g\, : \, a \in G_F \big\} $ is the \emph{$ G_F $-orbit} of $ g $. The conditions~\eqref{eq:V_on_surjective} and~\eqref{eq:V_on_injective} guarantee minimality of $ V_\on $ with flavor being the only remaining redundancy of $ V_\on $, which is an unavoidable consequence of using flavor-covariant operators. We have $ (V_\off/\! \sim) \cong V_\on / G_F $. Even the on-shell coupling space over-parameterizes physics, which is the source of various ambiguities in the RG.

\subsection{RG ambiguities and divergences} \label{sec:RG_divergence}
A central tenet of the RG is invariance of renormalized Green's functions w.r.t. variations of the renormalization scale. This can be cast as the Callan--Symanzik equation for the renormalized \emph{vacuum functional} $ W_\off[\lambda, \scJ] $ of the off-shell formulation:
	\begin{equation} \label{eq:CS}
	\left(\mu \dfrac{\partial }{\partial \mu} + \beta_\off^i \partial_i + \int \dd^d x \, \scJ_I(x) \gamma\ud{\off\, I}{J} \dfrac{\delta}{\delta \scJ_J(x)}  \right) \!W_\off[\lambda, \scJ] = 0.
	\end{equation}
The equation describes the flow of finite connected $ n $-point functions and was long understood to imply that the RG functions would have to be finite to ensure this. Ref.~\cite{Herren:2021yur} points out that there are unphysical directions in theory space---the directions generated by flavor rotations---and this allows for a particular class of divergences in the RG functions.

To appreciate this point, consider the action of the flavor group $ G_F $ at the infinitesimal level. If $ \alpha \in \mathfrak{g}_F $ is an element in the Lie algebra of $ G_F $, its induced action on the field is 
	\begin{equation}
	\delta_\alpha \eta^I = -(\alpha \cdot \eta)^I =  - \alpha\ud{I}{J} \eta^J,
	\end{equation}
where the anti-Hermitian matrix $ \alpha\ud{I}{J} $ is the suitable representation of $ \alpha $. Flavor invariance of the off-shell Lagrangian~\eqref{eq:Lag_off-shell_w/_source} dictates a simultaneous transformation of the couplings and sources: 
	\begin{align}
	\delta_\alpha \scJ_I &= -(\alpha \cdot \scJ)_I =  \scJ_J \alpha\ud{J}{I}, \\
	\delta_\alpha \lambda^i &=  -(\alpha\cdot \lambda)^i, 
	\end{align}
where the action $ (\alpha\cdot \lambda)^i $ on the couplings is informed by the fields appearing in the operator $ O_i $. The bare couplings and the source transform as the renormalized quantities, and all counterterms are flavor-covariant functions of the renormalized couplings. The vacuum functional is invariant under flavor rotations,\footnote{With possible exceptions due to chiral flavor anomalies~\cite{Keren-Zur:2014sva}. One would expect the treatment of such anomalies to result in a non-linear realization of the topological $ \theta $-term couplings under the flavor group.} resulting in the Ward identity 
	\begin{equation} \label{eq:flavor_WI}
	\left( - (\alpha \cdot \lambda)^i \partial_i + \int \dd^d x \, \scJ_I(x) \alpha\ud{I}{J} \dfrac{\delta}{\delta \scJ_J(x)}  \right) \! W_\off[\lambda, \scJ] = 0, \qquad \alpha \in \mathfrak{g}_F,
	\end{equation}
associated with the flavor symmetry. 

The flavor Ward identity allows for the possibility that even divergent RG functions in~\eqref{eq:CS} are compatible with finite flow of the renormalized Green's functions, as long as the divergences mimic the structure of~\eqref{eq:flavor_WI}. This motivates a relaxation of the requirement that RG functions must be finite; rather, it is sufficient that their $ \epsilon $-poles satisfy~\cite{Herren:2021yur} 
	\begin{equation} \label{eq:rg-finiteness}
	\gamma_{\off,n} \in \mathfrak{g}_F \andeq \beta_{\off,n}^i = - (\gamma_n \cdot \lambda)^i,\qquad \forall n \geq 1.
	\end{equation}
Here we use the notation $ \gamma_{n} \in \mathfrak{g}_F $ to indicate that there is an element of the Lie algebra that equals $ \gamma_n $ in the representation of the fields under the flavor group. RG functions satisfying these relations are said to be \emph{RG-finite}.

It is noteworthy that observables are built from flavor invariants of the theory. Hence,
	\begin{equation} \label{eq:observables_WI}
	0 = (\alpha \cdot g)^a \partial_a \mathcal{O}(g, \mu) = (\alpha \cdot \lambda)^i \partial_i \mathcal{O}(\lambda, \mu), \qquad \qquad \alpha \in \mathfrak{g}_F.
	\end{equation}
The condition of RG-finiteness, therefore, also guarantees a finite flow of observables. The same argument about flavor rotations being unphysical holds in the on-shell formulation, allowing also for RG-finite divergences (and flavor ambiguities) of the on-shell \befs. To entirely avoid this issue one should adopt a strict definition of what constitutes physical couplings, as commented on in Note~\ref{note:phys_befs}.

\begin{note}[note:phys_befs]{Physical parameters}
There is a more stringent interpretation of what constitutes physical couplings than what we have adopted at this point, which somewhat avoids the redundancy of the flavor group. One could argue that one should use the freedom of the flavor rotations to fix a ``gauge'' for the coupling tensors to write them in terms of a smaller set of physical parameters. For instance, one might write the nine fermion masses and four CKM parameters instead of the three $ 3\times 3 $ SM Yukawa matrices. Such a parametrization can be enforced with a unitary field transformation and would presumably ensure that there is no remnant redundancy in the RG and that the \befs of the physical parameters would be finite. Equivalently, one could employ a formulation in terms of basis invariants constructed from the flavored couplings to avoid specifying a gauge (see e.g.~\cite{Bento:2023owf} and references therein). We adopt a geometric view of this matter in Section~\ref{sec:geometry}.  
\end{note}

Divergences in RG functions are not just an academic possibility; they were encountered in the first calculation of three-loop RG functions in the SM~\cite{Bednyakov:2014pia} and 2HDM~\cite{Herren:2017uxn}. The authors realized that the two-point Green's functions, used to fix the wave-function renormalization factor in the combination $ Z^\dagger Z $, allow for tuning $ Z $ with an extra unitary factor (without changing renormalization scheme or finite Green's functions) by which the divergences could be removed. The freedom to change the renormalization constants and, thereby, the finite part of the RG functions was originally discussed in the context of the Local RG~\cite{Jack:1990eb,Fortin:2012hn,Jack:2013sha} and understood to be a consequence of the freedom to perform flavor rotations along the RG flow.  

Finite flavor rotations transform the renormalized couplings. If instead, we choose a divergent rotation $ U(\lambda) \in G_F $ of the bare theory, such that $ U= \mathds{1} + \ord{1/\epsilon} $, we can change the counterterms without a change in renormalized couplings.\footnote{In practice one would let $ U(\lambda) = e^{- \omega(\lambda)} $, with $ \omega(\lambda) \in \mathfrak{g}_F $ being a series in $ 1/\epsilon $ with no finite part, and truncate the expansion of the exponential at an appropriate order in loop-order and $ \epsilon $ poles.}
Letting $ \eta^I \to U\ud{I}{J} \eta^J $ leads to the transformation
	\begin{equation} \label{eq:ct_arbitrariness}
	Z\ud{I}{J} \longrightarrow U\ud{I}{K} Z\ud{K}{J}, \qquad \lambda_{\mathrm{b}}^i \longrightarrow  (U\cdot \lambda_{\mathrm{b}})^i 
	\end{equation}
of the wave-function renormalization and the bare couplings (coupling counterterms) of~\eqref{eq:Lag_off-shell_w/_source}. The transformation changes the counterterms without changing the renormalized couplings, and one concludes that the counterterms are not uniquely determined by the choice of renormalization scheme. As long as $ U(\lambda) $ is a flavor-covariant function of the renormalized couplings, the new counterterms will retain the covariance of the original counterterms. 

It was demonstrated in~\cite{Herren:2021yur} that the change in counterterms induced by $ U(\lambda) \in G_F $ leads to a change in RG functions of the form 
	\begin{equation} \label{eq:RG_ct_ambiguity}
	\gamma_\off \longrightarrow \gamma_\off + \Delta \gamma, \qquad \beta_\off^i \longrightarrow  \beta_\off^i - (\Delta \gamma\cdot  \lambda)^i, 
	\end{equation}
where $ \Delta \gamma(\lambda) \in \mathfrak{g}_F $. In fact, it is possible to choose $ U(\lambda) $ to obtain any covariant $ \Delta \gamma(\lambda) \in \mathfrak{g}_F $ that can be constructed from polynomials of the available couplings. The change in RG functions is guaranteed to preserve the finite flow of the Green's functions~\eqref{eq:CS} as a consequence of the flavor Ward identity~\eqref{eq:flavor_WI}. While RG-finiteness~\eqref{eq:rg-finiteness} is invariant w.r.t. this change in RG functions, it is always possible to recover finite RG functions with a suitable $ U(\lambda) $. The finite part of the RG functions is nevertheless ambiguous too. There exist special \emph{flavor-improved} versions of the RG functions, $ B^i $ and $ \Gamma $ (\bef and field anomalous dimension), that are unambiguous and more ``physical'' in that it is the vanishing of $ B^i $ that signals a conformal field theory~\cite{Fortin:2012hn,Jack:2013sha,Baume:2014rla,Herren:2021yur}. These flavor-improved RG functions are well-defined in the off-shell formulation of the EFT.

The ambiguity~\eqref{eq:RG_ct_ambiguity} from the counterterm choice propagates to the on-shell \befs as well. 
Applying the mapping formula~\eqref{eq:mapping_on/off-shell_bef} yields 
	\begin{equation} \label{eq:beta_on_ct_ambiguity}
	\beta_\on^a(g) \longrightarrow  \beta_\on^a(g) - (\Delta \gamma(\lambda(g)) \cdot g)^a,
	\end{equation}
as a consequence of flavor covariance in the infinitesimal form, which yields
	\begin{equation}
	\alpha\cdot g(\lambda) = (\alpha\cdot \lambda)^i \partial_i g(\lambda), \qquad  \alpha \in \mathfrak{g}_F . 
	\end{equation}
The arbitrariness of the on-shell \befs also manifests as the action of a Lie algebra element on the couplings and corresponds to an extra component of the flow in the direction of a flavor rotation.

We have now seen how the counterterm ambiguity can be used to change the RG functions, even as the renormalized Green's functions and the RG flow of Green's functions and observables remain unchanged. It can also be used to introduce or remove unphysical $ \epsilon $-poles in the RG functions. This ambiguity warrants close scrutiny when we interpret spurious poles in EFT RG functions. Before getting to that, we have identified another source of ambiguities in the \befs of the truncated on-shell formulation, which we briefly examine.

\subsection{Non-uniqueness of the projection from off- to on-shell} \label{sec:projection_nonuniqueness}
We observe that the projection $ g_\Xi(\lambda) = \lambda|_{\Xi(\lambda)} $ is not unique even after fixing the on-shell subspace $ V_\on \subseteq V_\off $ with an on-shell basis.\footnote{The subscript $ \Xi $ in the on-shell projection defined in~\eqref{eq:implicit_Xi_def} is used here to emphasize that this is the particular projection defined by using redefinition $ \Xi $.} If $ \Xi^I(\eta, \lambda) $ is a field redefinition that takes the off-shell Lagrangian to the on-shell form as in~\eqref{eq:implicit_Xi_def}, then for any covariant $ U(\lambda)\in G_F $,
	\begin{equation} \label{eq:field_redef_ambiguity}
	\Xi^{I}_U(\eta, \lambda) = [U^{\eminus 1}(\lambda)] \ud{I}{J} \Xi^{J}(\eta, \lambda)
	\end{equation}
is an equally valid redefinition for bringing the Lagrangian to the on-shell form. However, the new field redefinition endows a new projection 
	\begin{equation} \label{eq:g_xi_prime}
	g_{\Xi_U}(\lambda) = \lambda|_{\Xi_U} = U(\lambda) \cdot g_\Xi(\lambda).
	\end{equation}
Clearly, this implies an underlying ambiguity in the choice of map $ g_\Xi:\, V_\off \to V_\on $, which we used to describe the mapping from the off- to the on-shell formulation. In principle, it is even possible to choose a $ \Xi $ such that $ g_\Xi|_{V_\on} \neq \mathrm{id}_{V_\on} $ (the restriction of $ g_\Xi $ to $ V_\on \subseteq V_\off $ is not the identity function). Since the flavor group action closes in $ V_\on $, the change of field redefinition~\eqref{eq:field_redef_ambiguity} does not change the subspace $ V_\on $, nor the on-shell basis.  
We denote by $ \lambda_{\Xi}:\, V_\on \to V_\off $ a right-inverse to $ g_\Xi $~\eqref{eq:g_xi_prime}; that is, $ g_\Xi \circ \lambda_\Xi = \mathrm{id}_{V_\on} $. A corresponding right inverse $ \lambda_{\Xi_U} $ of $ g_{\Xi_U} $ is given by 
	\begin{equation}
	\lambda_{\Xi_U}(g) = U^{\eminus 1}\!\big( \lambda_{\Xi}(g)\big) \cdot \lambda_{\Xi}(g),
	\end{equation}
which may be verified using covariance of the projection.\footnote{Covariance implies that $ U(a\cdot \lambda) = a\, U(\lambda) \, a^{\eminus 1} $ and $ g_\Xi(a\cdot \lambda) = a\cdot g_\Xi(\lambda) $ for $ a \in G_F$.} 

Interestingly, different choices of maps $ g_\Xi $ to the on-shell basis result in different on-shell \befs~\eqref{eq:mapping_on/off-shell_bef} defined via the mapping from the off-shell \befs. The definition of the on-shell \befs with the mapping induced by $ \Xi_U $ yields 
	\begin{equation} \label{eq:beta_Xi_U_1}
	\begin{split}
	\beta^{a}_{\Xi_U}(g) &= \beta_\off^i(\lambda) \partial_i g_{\Xi_U}^a(\lambda) \bigg|_{\lambda= \lambda_{\Xi_U}(g)} \\
	&= \beta_\off^i(\lambda) \Big[ \big(U(\lambda) \cdot \partial_i g_{\Xi}(\lambda) \big)^{a} + \big(\partial_i U(\lambda) U^{\eminus 1}(\lambda) \cdot g_{\Xi_U}(\lambda) \big)^{a} \Big] \bigg|_{\lambda= \lambda_{\Xi_U}(g)}.
	\end{split}
	\end{equation}
This expression greatly simplifies using covariance of the various functions. In particular, the derivatives satisfy
	\begin{equation}
	\begin{aligned}
	\dfrac{\partial}{\partial \lambda^{\prime i}} g_\Xi(\lambda') \bigg|_{\lambda' = a\cdot \lambda} &= (a^{\eminus 1})\ud{j}{i}\, a\cdot \partial_j g_{\Xi}(\lambda),\\ 
	\dfrac{\partial}{\partial \lambda^{\prime i}} U(\lambda') \bigg|_{\lambda' = a\cdot \lambda} &= (a^{\eminus 1})\ud{j}{i}\, a\, \partial_j U(\lambda)\, a^{\eminus1},
	\end{aligned}
	\qquad \qquad a \in G_F,
	\end{equation}
whereas the off-shell \bef satisfies $ \beta^i(a\cdot \lambda) = a\ud{i}{j} \beta^j(\lambda) $.
With these observations~\eqref{eq:beta_Xi_U_1} reduces to 
	\begin{equation} \label{eq:beta_Xi_U_2}
	\beta^{a}_{\Xi_U}(g) = \beta_{\Xi}^a(g) + \big( \beta_\off^{i}(\lambda) \big[U^{\eminus 1}(\lambda) \partial_i U(\lambda) \big] \cdot g \big)^{\!a} \bigg|_{\lambda = \lambda_{\Xi}(g)}   
	\end{equation}
We recognize $ U^{\eminus 1} \partial_i U $ as the pullback of the Maurer--Cartan form on $ V_\off $ using $ U $ and conclude that $ \beta_\off^{i}(\lambda) \big[U^{\eminus 1}(\lambda) \partial_i U(\lambda) \big] $ is an element of the flavor group Lie algebra $ \mathfrak{g}_F $. The second term of~\eqref{eq:beta_Xi_U_2} is the discrepancy from the on-shell \bef as calculated by the original field redefinition $ \Xi $.

We observe that obtaining on-shell \befs by embedding the on-shell basis in the off-shell framework, calculating the off-shell \befs, and then mapping back to the on-shell basis provides another source of ambiguities, which to our knowledge has not been remarked on previously. Notably, the ambiguity is again unphysical as it generates a flow in the direction of a flavor transformation similar to the ambiguity~\eqref{eq:beta_on_ct_ambiguity} arising from the choice of counterterms. The ambiguity in the on-shell \befs from the second term in~\eqref{eq:beta_Xi_U_2} is likely to start only at the three-loop order and certainly not before two-loop order: $ \beta^i(\lambda) $ is a one-loop object, and non-trivial contributions to $ U(\lambda) $ require non-Hermitian, covariant contractions of the couplings (understood to be elements of $ \mathfrak{g}_F $). Such combinations certainly happen at two-loop order in EFTs (cf. Section~\ref{sec:nuSMEFT}), but it is unclear if there exist EFTs where they can be engineered at one-loop order. 

We wish to stress that this ambiguity in the on-shell \befs~\eqref{eq:beta_Xi_U_2} is not as artificial as it might look at first glance. One might reasonably expect practitioners to always choose the trivial embedding $ \lambda_\Xi(g) = g $, but this requirement does not uniquely fix the projection $ g_\Xi $. One should simply choose $ U(\lambda) $ subject to the constraint $ U(\lambda_\Xi(g)) = e $, where $ e \in G_F$ is the identity element. In other words, if we can find a function $ U(\lambda) $ that is non-trivial only when some ``redundant'' off-shell couplings are non-vanishing, then $ g_{\Xi} $ is ambiguous even after fixing $ \lambda_\Xi(g) = g $.

The ambiguity~\eqref{eq:beta_Xi_U_2} exists for renormalizable theories ($ V_\on = V_\off $) too. An extraordinarily mischievous practitioner might decide to perform a redefinition $ \Xi^{I}_U(\eta, \lambda) = \big(U(\lambda) \cdot \eta \big)^I $ (compared to the usual, reasonable choice $ \Xi^I (\eta, \lambda) = \eta^I $, i.e., performing no redefinitions), which would change the \befs. This corresponds to rotating the theory with a coupling-dependent transformation before calculating the \befs and rotating back again. Although this idea is both highly unusual and impractical, it is in principle not prohibited.

\section{On-Shell Formulation with Non-Minimal Source Terms} \label{sec:NMS}
We return to the question of field redefinitions and equivalence between theories. This time we include all source terms to ensure that all off-shell Green's functions are faithfully reproduced in the full on-shell formulation of the theory.

\subsection{On-shell vacuum functional}
The vacuum functional of the off-shell Lagrangian~\eqref{eq:Lag_off-shell} is
	\begin{equation} \label{eq:vf_off-shell}
	W'_\off[\lambda, \scJ^\off] = \eminus i \log \! \int \mathcal{D}\eta\, \exp \! \left[ i \!\left(S_\off[\eta,\lambda] + \!\int \dd^d x \, \scJ^\off_I(x) \eta^I(x)  \right) \right],
	\end{equation}
where $ S_\off = \!\int \, \dd^d x \L_\off $ is the associated action. The role of the fields $ \eta^I $ as integration variables is manifest in the path integral. The transformation $ \eta^I \to \Xi^I\! (\eta, \lambda) $~\eqref{eq:implicit_Xi_def}, which brings the Lagrangian (action) to the on-shell basis, leaves the vacuum functional invariant:
	\begin{equation}
	W'_\off[\lambda, \scJ^\off] = \eminus i \log \! \int \mathcal{D}\eta\, \exp \! \left[ i \!\left(S_\on[\eta, g(\lambda)] + \! \int \dd^d x\, \scJ_I^\off(x)  \Xi^I\! (\eta(x),\lambda) \right) \right],
	\end{equation}
having used that perturbative field redefinitions do not change the measure of the path integral in dimensional regularization~\cite{Criado:2018sdb}.
While the action is now in its on-shell form, the trade-off is the introduction of \emph{non-minimal source terms}---source terms other than $ \scJ^\off_I \eta^I $.\footnote{One might say that $ \scJ^\off_I $ now sources the composite operator $ \Xi^I(\eta,\, \lambda) $; however, it is useful for our purposes to think of the source being part of multiple distinct terms instead, which all renormalize independently.} 

The NMSTs are parametrized by setting 
	\begin{equation} \label{eq:field_redefinition_parametrization}
	\Xi^I\! (\eta,\lambda)  = A\ud{I}{J}(\lambda) \left(\eta^J + r^\alpha(\lambda) Q_\alpha^J(\eta) \right), 
	\end{equation}
where $ Q_\alpha^I(\eta) $ is a basis of compound operators that transform like $ \eta^I $ (under symmetries) up to the required order in the EFT expansion. A factor $ A\ud{I}{J}(\lambda) $ is generally needed to ensure that the canonical normalization of $ S_\mathrm{kin} $ is maintained after the transformation. It also allows for the unitary flavor rotations discussed in Section~\ref{sec:flavor_group}. This factor is simply absorbed into a rescaling of the sources. The operators $ Q_\alpha^I $ are in one-to-one correspondence with the ``redundant'' operators eliminated when going from $ S_\off $ to $ S_\on $. With $ V_\mathrm{red} = \mathrm{span}\big(\{Q_\alpha^I\}\big) $ being the vector space of these redundant NMSTs, we let $ r = r^\alpha Q_\alpha^I \in V_\mathrm{red} $. It follows that 
	\begin{equation}
	\dim V_\off = \dim V_\on + \dim V_\mathrm{red}.
	\end{equation} 
This leads us to the interpretation of $ r^\alpha(\lambda) $ as a set of ``redundant'' couplings, such that the change from off-shell to on-shell theory does not change the number of independent couplings.  

The equivalent on-shell version of the vacuum functional~\eqref{eq:vf_off-shell} is 
	\begin{equation} \label{vf_on-shell}
	W'_\on[g,\, r,\, \scJ^\on] = \eminus i \log \! \int \mathcal{D}\eta\, \exp \! \left[ i \!\left(S_\on[\eta, g] +\! \int \dd^d x\, \scJ^\on_I(x) \big[\eta^I + r^\alpha Q^I_\alpha(\eta) \big]  \right) \right].
	\end{equation}
We refer to this form as the \emph{full on-shell formulation} of the theory. The field redefinition $ \Xi^I(\eta, \lambda) $ establishes a map 
	\begin{equation} \label{eq:invertible_off-on_map}
	\big(g^{a},\, r^{\alpha},\, \scJ_I^\on\big) = \big( g^{a}(\lambda),\, r^{\alpha}(\lambda),\, \scJ_I^\on(\lambda, \scJ^\off) \big).
	\end{equation}
composed of the projection $ g(\lambda) $~\eqref{eq:implicit_Xi_def} to the on-shell coupling space and the parametrization~\eqref{eq:field_redefinition_parametrization}. 
The map is invertible---one needs simply find a field redefinition to remove all NMSTs~\cite{Anselmi:2012aq}---so it allows for a unique inverse mapping from on-shell to off-shell couplings: 
	\begin{equation}
	(\lambda^{i},\, \scJ_I^\off) = \big(\lambda^{i}(g, r),\, \scJ_I^\off(g,r,\scJ^\on) \big).
	\end{equation}
This is in contrast to the arbitrary choice of right inverse $ \lambda(g) $ used in Section~\ref{sec:prelims} for mapping $ \L_\on $ to $ \L_\off $. In practice, the effect of the mapping is to move redundant couplings between off-shell operators and new NMSTs without any change in off-shell Green's functions. The reader may reasonably object to the use of ``on-shell formulation'' when referring to~\eqref{vf_on-shell}, as it contains exactly the same information as the off-shell formulation~\eqref{eq:vf_off-shell}. We use this name to emphasize that it is a superset of the truncated on-shell formulation, often referred to as an on-shell EFT basis. Additionally, the full on-shell formulation clearly separates on-shell couplings from redundant ones---even under the RG, as we shall see in due course---which can be omitted in many calculations. 

\begin{figure}
	\centering
	\includegraphics{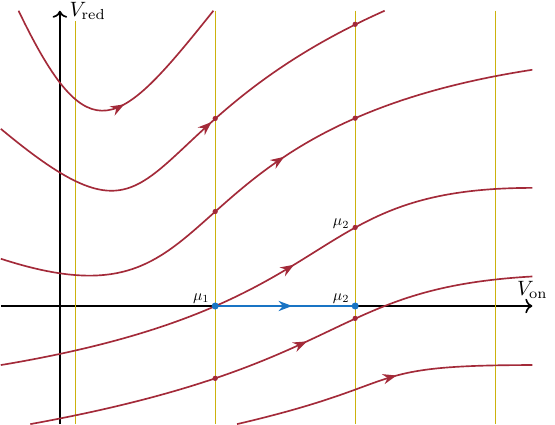}
	\caption{Illustration of the RG flow in full on-shell formulation. The horizontal axis indicates the (multi-dimensional) space of on-shell couplings, while the redundant couplings of the NMSTs are on the vertical axis. The yellow lines indicate physically equivalent surfaces, which in this formulation are all points with identical (modulo the $ G_F $ action) on-shell couplings (by contrast to Fig.~\ref{fig:off-shell_flow}). The red lines with arrows indicate the flow of the couplings under RG. Under RG flow, all points on a yellow line at scale $ \mu_1 $ flow to points on another yellow line at scale $ \mu_2 $. This is a consequence of physics being fully specified by the on-shell couplings. The blue arrow indicates the flow of on-shell couplings between $ \mu_1 $ and $ \mu_2 $.}
	\label{fig:on-shell_flow}
	\end{figure}

Physical observables, derived from $ S $-matrix elements, are independent of the exact source terms used to interpolate a particle. It is well-established that the non-minimal piece of the source term may be dropped ($ r \to 0 $) without changing the $ S $-matrix elements~\cite{Arzt:1993gz,Criado:2018sdb}. One needs to keep $ r \neq 0 $ only to maintain off-shell equality of the vacuum functional. We consider the product space $ V_\on \times V_\mathrm{red} $ to be the theory space in the full on-shell formulation. The equivalence classes of physically identical theories are given by
	\begin{equation} \label{eq:on-shell_equiv_class}
	[(g,r)] = \big\{ (g',r')\,:\, g' \in G_F \cdot g,\, r' \in V_\mathrm{red} \big\}, \qquad (g,r) \in V_\on \times V_\mathrm{red}.
	\end{equation}
All physics is determined by the $ V_\on $ factor in the product space, although there is still the remaining redundancy due to flavor transformations. This is much simpler than the non-linear surfaces~\eqref{eq:off-shell_equiv_classes} of physical equivalence in the off-shell formulation. The equivalence classes are illustrated as yellow lines in the full on-shell parameter space shown in Fig.~\ref{fig:on-shell_flow}.\footnote{We caution the reader that the intersections of the equivalence classes with $ V_\on $ are $ G_F $-orbits rather than points, as suggested by this figure (cf. also Fig.~\ref{fig:flow_projection}). In principle, the redundant couplings can influence the associated running in the $ V_\on $ direction, but only to the extent that they can generate non-physical flavor rotations in $ V_\on $.}

Given the off-shell equivalence between the two formulations, we have that $ V_\on \times V_\red $ is isomorphic to $ V_\off $ and shares its over-parametrization of physics---as indicated by referring to $ r^\alpha $ as redundant couplings. Working in the $ V_\on \times V_\red $ space most likely over-complicates many calculations, and we may want to revert to the truncated on-shell formulation at some point and perform the RG evolution in that formulation. A projection $ \psi: V_\on \times V_\red \to V_\on $ is needed for this purpose, which seems trivial on its face. Nevertheless, $ \psi $ is ambiguous just as was the projection $ g: V_\off \to V_\on $. The most general projection preserving the equivalence class~\eqref{eq:on-shell_equiv_class} (in the sense that $ \big(\psi(g,r),\, 0\big) \in [(g,\,r)] $) is  
	\begin{equation}
	\psi(g, r) = U(g, r) \cdot g
	\end{equation} 
for some $ G_F $-covariant function $ U: V_\on \times V_\red \to G_F $. This is simply an expression of the remaining ambiguity inherent in $ V_\on $: the same ambiguity will naturally be reflected in the \befs too.

\subsection{Renormalization and RG functions}
We saw in Section~\ref{sec:cts_and_rgs_off-shell} how the off-shell Lagrangian was renormalized with bare couplings and sources. We may define the renormalized vacuum functional of the off-shell basis through 
	\begin{equation} \label{eq:W_renormalization_off-shell}
	W_\off[\lambda, \scJ^\off] \equiv W'_\off[\lambda_\mathrm{b}, \scJ^\off_\mathrm{b}] .
	\end{equation}
The bare couplings and sources are functions of their renormalized counterparts and give rise to counterterms as per the parametrizations~\eqref{eq:bare_cts} and~\eqref{eq:Z_parameterization}. The core tenet of renormalization is to introduce a counterterm for any symmetry-allowed operator of the fields and sources (up to the working order in the EFT power counting). The off-shell Lagrangian consists of a full basis of the field operators, and any hypothetical counterterms involving NMSTs can be shifted into counterterms for the off-shell couplings with a field redefinition. Thus, the renormalized vacuum functional~\eqref{eq:W_renormalization_off-shell} contains a sufficient set of counterterms to fully renormalize the off-shell Green's functions. 

The renormalization of the on-shell vacuum functional mirrors the off-shell version:\footnote{An alternative approach to renormalize the vacuum functional after a field redefinition was proposed in~\cite{Anselmi:2012aq}. The idea is to introduce sources for all composite operators (reminiscent of the spacetime-dependent couplings introduced in the local RG) and redefine these sources to remove the NMSTs.}
	\begin{equation}
	W_\on[g,\, r,\, \scJ^\on] \equiv W'_\on[g_\mathrm{b},\, r_\mathrm{b},\, \scJ^\on_\mathrm{b}].
	\end{equation}
The bare couplings are now functions of both on-shell and redundant renormalized couplings. The counterterms do not span the full operator basis, but any hypothetical counterterm for an operator not in the on-shell basis can be shifted to counterterms for the NMSTs---the $ \delta r^\alpha $ counterterms---instead.\footnote{One might speculate that additional counterterms involving multiple sources in the same operator might be needed, as they are allowed on dimensional grounds. We expect that such terms can be eliminated by shifting the field $ \eta \to \eta + f(\eta, \scJ) $, for some function $ f(\eta, \scJ) $ of the field and source. A field redefinition of this type was employed in~\cite{Naterop:2024cfx}.} The invertible map~\eqref{eq:invertible_off-on_map} between off- and on-shell couplings can also be used for the bare couplings, preserving renormalization of off-shell Green's functions. 

Having identified a complete on-shell formulation, which includes the NMSTs and a fully renormalized vacuum functional, we can assert invariance of the bare couplings and sources and from there derive formulas for the \befs for on-shell and redundant couplings. The \befs are written as 
	\begin{equation}
	\beta_\on^a(g,r) \equiv \dfrac{\dd g^a}{\dd t}, \qquad \beta_\red^\alpha(g,r) \equiv \dfrac{\dd r^\alpha}{\dd t},
	\end{equation}
each with a loop and pole expansion similar to~\eqref{eq:beta_parameterization} for the off-shell couplings. The recursive formulas for the poles of the on-shell \befs in terms of counterterms [cf.~\eqref{eq:beta_formula_off-shell}] are
	\begin{align} 
	\beta_{\on,n}^a &= (\zeta - k_a) \delta_{n+1} g^a - \sum_{m=0}^{n-1} \big(\beta_{\on,m}^b \partial_b + \beta_{\red,m}^\beta \partial_\beta \big) \delta_{n-m} g^a, \label{eq:phys_beta-function_formula} \\
	\beta_{\red,n}^{\alpha} &= (\zeta - k_\alpha ) \delta_{n+1} r^\alpha - \sum_{m=0}^{n-1} \big(\beta_{\on,m}^b \partial_b + \beta_{\red,m}^\beta \partial_\beta \big) \delta_{n-m} r^\alpha, \label{eq:red_beta-function_formula}
	\end{align}
for $ n\geq 0 $, where $ \partial_\alpha \equiv \frac{\partial}{\partial r^\alpha} $ in analogy with the other couplings. The field-anomalous dimension in the on-shell formulation is 
	\begin{equation} \label{eq:on-shell_gamma_formula}
	\gamma_{n}^\on = \zeta Z_{n+1} + \sum_{m=0}^{n-1} \left[ \big(\beta_{\on,m}^b \partial_b + \beta_{\red,m}^\beta \partial_\beta \big) Z_{n-m} - Z_{n-m} \gamma^\on_m \right]. 
	\end{equation}
The renormalization scale invariance of the bare vacuum functional, then implies the Callan--Symanzik equation 
	\begin{equation} \label{eq:CS_on-shell}
	\left( \mu \dfrac{\partial }{\partial \mu} + \beta_\on^a \partial_a + \beta_\red^\alpha \partial_\alpha + \int \dd^d x \, \scJ^\on_I(x) \gamma\ud{\on\,I}{J} \dfrac{\delta}{\delta \scJ^\on_J(x)}  \right) \!W[g,\, r, \scJ^\on] = 0
	\end{equation}
for the renormalized vacuum functional, dictating the RG flow of the Green's functions.
The on-shell formulation with NMSTs is entirely equivalent to the off-shell formulation. Both parametrizations produce finite off-shell Green's functions and are equally valid renormalized descriptions of the same underlying theory. Based on this equivalence, we expect both sets of RG functions to generate finite RG flow; i.e., the full on-shell RG functions should be RG-finite. 

The coupling RG flow in the full on-shell coupling space $ V_\on \times V_\mathrm{red} $ is better behaved than the off-shell flow depicted in Fig.~\ref{fig:off-shell_flow}. The red lines with arrows in Fig.~\ref{fig:on-shell_flow} illustrate the RG flow in the full theory space that leaves all off-shell Green's functions invariant. Any point in the same physical equivalence class should flow to another shared equivalence class after the same RG time, which is indicated as the flow between red dots at scale $ \mu_1 $ and $ \mu_2 $. 

The truncated on-shell flow in $ V_\on $ (depicted with the blue arrow) can be determined by embedding $ V_\on $ in $ V_\on \times V_\red $, using the full on-shell \befs and projecting back [cf.~\eqref{eq:mapping_on/off-shell_bef}]. This construction has the same inherent ambiguities from the specific projection choice (including, presumably, the choice of right inverse) as the construction of the on-shell \befs through $ \beta_\off $. In the simplest possible version with $ \psi(g,r)= g $ and the right inverse $ \imath: g\mapsto (g,0) $, the \bef on $ V_\on $ is simply 
	\begin{equation}\label{eq:beta_on_from_full_on-shell}
	\beta_\on(g) = \beta_\on(g,r=0).
	\end{equation}
While this construction can be made (artificially) more convoluted for other choices of $ \psi, \imath $, the simplest version of the on-shell formulation with NMSTs factorizes the RG flow into an on-shell and a redundant direction, of which the former is independent of the redundant couplings.

That the redundant couplings have a purely unphysical impact on the running of the on-shell couplings also follows from the structure of the counterterms. The NMSTs involve field sources, so any divergence involving such a term should itself involve a source. The source counterterm (wave-function renormalization factor) and redundant counterterms of the NMSTs are sufficient to renormalize the redundant couplings. The counterterms of the on-shell couplings renormalize subdivergences and vacuum diagrams not involving any external source terms; one would expect them to be entirely independent of any source terms. An exception to this logic is that it is possible to introduce dependence of the redundant couplings to the on-shell counterterms, in the sense that $ \delta g^a(g,r) \neq \delta g^a(g,0) $, through a divergent flavor rotation $ U(g,r) \in G_F $ similar to the one in~\eqref{eq:ct_arbitrariness}. As discussed in Section~\ref{sec:RG_divergence}, the impact of such a rotation is to alter the RG functions to generate an unphysical flavor rotation along the RG flow. Even in this case, the impact of the redundant couplings on the running of the on-shell couplings is unphysical. 

As we have argued, the usual \befs from the truncated on-shell formulation, without NMSTs, do not describe the flow of any Green's functions. Nevertheless, the formula for the \befs of the on-shell couplings~\eqref{eq:phys_beta-function_formula} and the field anomalous dimensions~\eqref{eq:on-shell_gamma_formula} resembles what one would derive in the absence of redundant couplings, with the notable exception of the $ \beta_{m}^\beta \partial_\beta  $-terms. The finite part of the on-shell \bef~\eqref{eq:beta_on_from_full_on-shell} does not involve any counterterms for the redundant couplings. While the $ \delta_1 g^a $ and $ Z_1 $ counterterms may have dependence on $ r $, we can certainly compute the finite RG functions at $ r^\alpha =0 $, without ever thinking about NMSTs, and obtain a suitable on-shell running. This reassures us that previous literature results are correct when calculated in the truncated on-shell formulation, even when omitting all effects of the NMSTs/redundant couplings. The same cannot be said for the divergent part of the RG functions, where neglecting the $ \beta_{m}^\beta \partial_\beta $-terms in~\eqref{eq:phys_beta-function_formula} and, in particular,~\eqref{eq:on-shell_gamma_formula} will lead to incorrect results. From a practical standpoint, even the erroneous divergent parts of $ \beta_\on $ obtained this way would generate flow in an unphysical direction generated by a flavor rotation and can be safely ignored.

We proceed to reexamine some recent examples from literature that have produced divergent RG functions when using a truncated on-shell formulation and demonstrate how the situation is remedied in the full on-shell formulation.

\section{Examples} \label{sec:examples}
To illustrate the impact of NMSTs on RG functions in the on-shell formulation, we consider two recent examples from the literature~\cite{Manohar:2024xbh,Zhang:2025ywe}. Both of these calculations encountered divergent RG functions in the truncated on-shell formulation; the second example from the $ \nu $SMEFT even reported seemingly spurious divergences in Yukawa coupling \befs. Conveniently, both works present enough details for us to identify the missing contributions from the NMSTs without redoing the full two-loop calculation.

\subsection{Dimension-six scalar $ \mathrm{O}(n) $ model}
Ref.~\cite{Manohar:2024xbh} observed that the field anomalous dimension of a scalar $ \mathrm{O}(n) $ model in a truncated on-shell formulation (without NMSTs) was divergent at two-loop order. We are now in a position to appreciate the origin of this puzzling result. In the notation of~\cite{Manohar:2024xbh}, the dimension-six off-shell EFT Lagrangian is 
	\begin{multline} \label{eq:On_Lagrangian_off-shell}
	\L_\off= \tfrac{1}{2} \partial_\mu \phi  \cdot \partial^\mu \phi- \tfrac{1}{2} m_\off^2 \phi\cdot \phi - \tfrac{1}{4}\lambda_\off (\phi\cdot \phi)^2 + C^\off_4 (\partial_\mu \phi  \cdot \partial^\mu \phi) (\phi\cdot \phi) + C^\off_6 (\phi\cdot \phi)^3 \\
	+ D^\off_4 (\phi\cdot \partial_\mu \phi)^2 + D^\off_2 (\partial^2 \phi \cdot \partial^2 \phi) + \scJ^\off \cdot \phi ,
	\end{multline}
where $ \phi_I $ is a real scalar field with $ n $ components [the `$ \cdot $' indicates a scalar product in the fundamental representation of $ \mathrm{O}(n) $]. Since the $ \mathrm{O}(n) $ symmetry is exact, all couplings are flavor singlets and we do not anticipate any ambiguity (or divergences) related to rotations in flavor space. We have added sub-/superscript ``off'' to the couplings to discriminate them from the equivalent couplings in the on-shell formulation. The Lagrangian~\eqref{eq:On_Lagrangian_off-shell} constitutes a full off-shell basis for the theory (including a source for the field). The operators associated with the couplings $ D^\mathrm{off}_{2,4} $ are considered redundant w.r.t. our eventual choice of on-shell basis, meaning that their coefficients can be removed by field redefinitions. In the concise notation we have used previously, the off-shell couplings of the theory are
	\begin{equation}
	\lambda^i = \big(m^2_\off,\,  \lambda_\off,\, C^\off_4,\, C^\off_6,\, D^\off_2,\, D^\off_4 \big).
	\end{equation}
If we shift to a full on-shell formulation of the theory, we obtain the on-shell Lagrangian\footnote{Performing an additional field redefinition $ \phi \to \phi + r_2 \scJ^\on $ can further remove the $ r_2 $ NMST in favor of a $ r_2 (\scJ^\on \cdot \scJ^\on) $ term (similar to~\cite{Naterop:2024cfx}). Conversely, this possibility indicates that we do not need to include the extra quadratic source counterterm, which is otherwise allowed on dimensional grounds.}
	\begin{multline} \label{eq:On_Lagrangian_on-shell}
	\L_\on= \tfrac{1}{2} \partial_\mu \phi  \cdot \partial^\mu \phi- \tfrac{1}{2} m^2_\on \phi\cdot \phi - \tfrac{1}{4}\lambda_\on (\phi\cdot \phi)^2 + C^\on_4 (\partial_\mu \phi  \cdot \partial^\mu \phi) (\phi\cdot \phi) + C^\on_6 (\phi\cdot \phi)^3 \\
	+ \scJ^\on\cdot (\phi +r_1 (\phi\cdot \phi)\phi + r_2 \partial^2 \phi).
	\end{multline}
with the on-shell and redundant couplings 
	\begin{equation}
	g^a = \big(m^2_\on,\,  \lambda_\on,\, C^\on_4,\, C^\on_6\big), \qquad r^\alpha = \big(r_1,\, r_2 \big).
	\end{equation}
The field redefinition required to go from off-shell to on-shell Lagrangians is 
	\begin{equation}
	\phi \longrightarrow (1- m_\off^2 D^\off_2) \phi - (\tfrac{1}{2} D^\off_4 + \lambda_\off D^\off_2 ) (\phi\cdot \phi) \phi  + D^\off_2 \partial^2 \phi + \ldots, 
	\end{equation}
neglecting higher-order terms in the EFT expansion. This indicates that the map between off-shell and full on-shell formulation is determined by 
	\begin{align}
	m_\on^2 & = m^2_\off - 2 m^4_\off D_2^\off, &
	\lambda_\on & = \lambda_\off - 2 m_\off^2 D_4^\off - 8 m^2_\off \lambda_\off D_2^\off, \nonumber \\
	C_4^\on &= C_4^\off - \tfrac{1}{2} D_4^\off, &
	C_6^\on &= C_6^\off + \tfrac{1}{2} \lambda_\off D_4^\off + \lambda_\off^2 D_2^\off \nonumber, \\
	r_1 &= -\tfrac{1}{2}D^\off_4 - \lambda_\off D_2^\off, &
	r_2 &= D_2^\off,  \nonumber \\
	\scJ^\on &= \scJ^\off (1 -m_\off^2 D_2^\off) \label{eq:scalar_ON_map}.
	\end{align}

Ref.~\cite{Manohar:2024xbh} found that the field anomalous dimension $ \gamma_\phi $ was finite when calculated in the off-shell formulation; however, in the truncated on-shell formulation (not accounting for the redundant couplings of the NMSTs) the authors found it to have a divergent (simple-pole) piece at two-loop order:
	\begin{equation}\label{eq:On_ADM_pole}
	\gamma^{(2)}_{1,\phi} \Big|_{\cancel{\sscript{NMST}}} = 2(n^2 -4) m_\on^2 \lambda_\on C_4^\on. 
	\end{equation}
As we will now argue, this divergence is spurious and is a consequence of not accounting for the redundant couplings of the on-shell formulation. We calculate, for simplicity, the on-shell running at a representative point $ r^\alpha = 0 $ for the physical equivalence class. The formula~\eqref{eq:on-shell_gamma_formula} for the simple pole of $ \gamma_\phi^{(2)} $ has a missing contribution involving $ \beta^{(1),\alpha}_{\red, 0}(g, r=0) $. Going from the one-loop off-shell counterterms provided in~\cite{Manohar:2024xbh}, we readily find (by mapping the bare couplings) that the counterterms of the redundant couplings are 
	\begin{equation}
	\delta^{(1)}_{1} \! r^\alpha(g, r=0) = \big((n -2) \lambda_\on C_4^\on , \, 0\big),
	\end{equation}
which implies the running 
	\begin{equation}
	\beta^{(1),r_1}_{\red, 0}(g, r=0) = 2 (n -2) \lambda_\on C_4^\on
	\end{equation}
of $ r_1 $. Of the redundant couplings, only $ r_1 $ runs at the representative point in the on-shell formulation. 

	\begin{figure}[t]
	\centering 
	\includegraphics*[width=.5\textwidth]{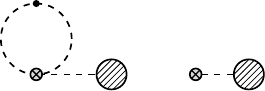}
	\caption{The divergence of the NMST of the scalar $ \mathrm{O}(n) $ theory on the left is renormalized with a counterterm of the ordinary source term (right). The field source $ \scJ $ is denoted by a cross on a gray background; the dot is a mass term; and the external scalar line connects with the rest of the Green's function, indicated with a hatched blob. 
	}
	\label{fig:scalar_source}
	\end{figure}

To determine the missing contribution to $ \gamma_\phi^{(2)} $ in the full on-shell formulation, all we need to compute is the $ r_1 $ contribution to $ Z_{1,\phi}^{(1)} $. We find that the wave-function renormalization of $ \phi $ receives the one-loop contribution 
	\begin{equation}
	Z^{(1)}_{1,\phi} \supset - (n+2) m^2_\on r_1 
	\end{equation}
from the diagram in Fig.~\ref{fig:scalar_source}.\footnote{Calculated here with the \texttt{Matchete} package~\cite{Fuentes-Martin:2022jrf} for convenience. The same goes for~\eqref{eq:Zl_contribution}. } Rather than calculating the contribution from the redundant couplings to the wave-function renormalization factor directly, one can also derive it from the off-shell counterterms presented in~\cite{Manohar:2024xbh} combined with the map~\eqref{eq:scalar_ON_map}.
The missing contribution to the first pole of the field anomalous dimension is 
	\begin{equation}
	\gamma^{(2)}_{1,\phi} \Big|_{\sscript{NMST}} = \beta^{(1),r_1}_{\red,0} \partial_{r_1} Z^{(1)}_{1,\phi} \Big|_{r=0} = - 2(n^2 -4) m_\on^2 \lambda_\on C_4^\on,
	\end{equation}
which is exactly the missing contribution needed to cancel~\eqref{eq:On_ADM_pole} and render the field anomalous dimension finite. We conclude that including the redundant couplings of the NMSTs in the on-shell formalism cures the spurious divergence of the field anomalous dimension arising from incomplete renormalization.

\subsection{Dimension-five $ \nu $SMEFT} \label{sec:nuSMEFT}
The recent two-loop calculation of the dimension-five RG functions in the $ \nu $SMEFT~\cite{Zhang:2025ywe} also encountered issues with divergent contributions in the on-shell basis, not only with the field-anomalous dimensions but, more troubling, with some \befs too. This theory is more complex than what we can readily examine in an example, so we single out one of the contributions and locate the source of the associated divergences. 

Part of the off-shell Lagrangian reads (following the notation of~\cite{Zhang:2025ywe})
	\begin{multline} \label{eq:nuSMEFT_off}
	\L_\off \supset \abs{D_\mu H}^2 + i\overline{\ell}_\LL \slashed D \ell_\LL + i \overline{N}_\RR \slashed \partial  N_\RR + \Big[ -\tfrac{1}{2} \overline{N}^c_\RR M_N^\off N_\RR - \overline{\ell}_\LL Y_\nu^\off \widetilde{H} N_\RR \\
	+ \overline{N}^c_\RR C_{HN}^\off N_\RR \abs{H}^2 + i \overline{\ell}_\LL \overleftarrow{\slashed D} G_{\ell H N 2}^\off \widetilde{H} N^c_\RR
	+\overline{\scJ}^\off_{\ell} \ell_\LL 
	\hc \Big]+ \ldots \,,
	\end{multline}
where $ H $ and $ \ell_\LL $ are $ \SU(2) $ doublets, and both fermions (left-handed leptons $ \ell $ and right-handed neutrinos $ N $) come in three flavors, making the relevant part of the flavor group $ G_F = \U(3)_\ell \times \U(3)_N \times \ldots $ The couplings $ M_N^\off $, $ C_{HN}^\off $, $ Y_\nu^\off $, and $ G_{\ell HN2}^\off $ are all $ 3\times 3 $ complex matrices in flavor space, the first two of which are symmetric. The coupling $ G_{\ell H N 2}^\off $ multiplies an EOM-type operator and is redundant. 

The equivalent on-shell formulation of the theory is given by 
	\begin{multline} \label{eq:nuSMEFT_on}
	\L_\on \supset \abs{D_\mu H}^2 + i\overline{\ell}_\LL \slashed D \ell_\LL + i \overline{N}_\RR \slashed \partial  N_\RR - \Big[\tfrac{1}{2} \overline{N}^c_\RR M_N^\on N_\RR + \overline{\ell}_\LL Y_\nu^\on \widetilde{H} N_\RR \\
	+ \overline{N}^c_\RR C_{HN}^\on N_\RR \abs{H}^2 
	+\overline{\scJ}^\on_{\ell} \big( \ell_\LL + r_\ell \widetilde{H} N^c_\RR \big) 
	\hc \Big]+ \ldots\,, 
	\end{multline}
where the coupling $ r_\ell $ of the NMST is redundant (and a $ 3\times 3$ complex matrix in flavor space). It is obtained by redefining the left-handed lepton field as
	\begin{equation}
	\ell_\LL \longrightarrow \ell_\LL + G_{\ell HN2}^{\off} \widetilde{H} N_\RR^c + \ldots 
	\end{equation}
in the off-shell Lagrangian~\eqref{eq:nuSMEFT_off}. The resulting map between the couplings is 
	\begin{align}
	M_N^\on &= M_N^\off, &
	Y_\nu^\on &= Y_\nu^\off, \nonumber \\
	C_{HN}^\on &= C_{HN}^\off - \tfrac{1}{2} \big(G_{\ell H N 2}^{\off\, \dagger} Y^\off_\nu + Y^{\off \, \intercal}_\nu G_{\ell H N 2}^{\off\, \ast} \big), &
	r_\ell &= G_{\ell H N 2}, \nonumber\\
	\scJ_\ell^\on &= \scJ_\ell^\off.  \label{eq:nuSMEFT_map}
	\end{align}
Ref.~\cite{Zhang:2025ywe} determined the divergent RG functions in the truncated on-shell formulation (without the redundant couplings of the NMSTs). 
The contributions from $ C_{HN}^\on $  to the simple poles are
	\begin{align}
	\gamma^{(2)}_{1,\ell} \Big|_{\cancel{\sscript{NMST}}} &\supset - Y_\nu^\on  \big( C_{H N}^{\on\, \dagger} M_N^\on + M^{\on \, \dagger}_N C_{ H N}^\on \big) Y_\nu^{\on\, \dagger}, \label{eq:ell_ADM_div}\\
	\beta^{(2),Y^\on_\nu}_{\on, 1} \Big|_{\cancel{\sscript{NMST}}} &\supset - Y_\nu^{\on} \big( C_{H N}^{\on\, \dagger} M_N^\on - M_N^{\on\, \dagger} C_{H N}^{\on} \big) Y_\nu^{\on\,\dagger } Y_\nu^\on \label{eq:beta_Ynu_div}
	\end{align}
for the $ \ell $ anomalous dimension and the Yukawa coupling \befs, respectively. At first this looks like a more serious issue with the RG flow, as compared to the $ \mathrm{O}(n) $ scalar example: here the \bef of an on-shell coupling exhibits an $ \epsilon $-pole. The astute reader may, however, recognize that the \bef divergence~\eqref{eq:beta_Ynu_div} takes the form of a flavor rotation along the RG flow. Hence, we can expect it to be consistent with RG finiteness~\eqref{eq:rg-finiteness}.\footnote{Originally, RG-finiteness is formulated as a combined condition on the coupling \befs and the field anomalous dimension, the latter of which comes out incorrect when using the truncated on-shell formulation. The flow generated by the \bef~\eqref{eq:beta_Ynu_div} nevertheless results in a finite flow of observables already without any corrections.} Curiously, the divergence of the field anomalous dimension~\eqref{eq:ell_ADM_div} is Hermitian, contrary to the requirement from RG-finiteness. This indicates that also in this case, the truncated on-shell formulation fails to generate a finite flow of the vacuum functional.

	\begin{figure}[t]
	\centering 
	\includegraphics*[width=.5\textwidth]{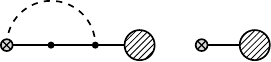}
	\caption{The divergence of the NMST of the $ \ell $ field in $ \nu $SMEFT on the left is renormalized with a counterterm of the ordinary source term (right). The field source $ \scJ $ is denoted with a cross on a gray background; the dots are renormalizable couplings (mass insertion and Yukawa interaction); and the external fermion line connects to the rest of the Green's function, indicated by a hatched blob. }
	\label{fig:fermion_source}
	\end{figure}

The RG functions receive an additional correction from the inclusion of the redundant coupling $ r_\ell $ to the on-shell formulation. Between the off-shell counterterms of~\cite{Zhang:2025ywe} and the map~\eqref{eq:nuSMEFT_map}, we find that the one-loop counterterm for the redundant coupling is 
	\begin{equation}
	\delta_{1}^{(1)} \! r_\ell = - Y_\nu^\on C_{HN}^{\on\, \dagger},
	\end{equation}
evaluated at $ r^\alpha =0 $. If the NMST is present, it necessitates a contribution to the wave-function renormalization to cancel the divergence of the diagram shown in Fig.~\ref{fig:fermion_source}. By direct calculation, we find
	\begin{equation} \label{eq:Zl_contribution}
	Z^{(1)}_{1,\ell} \supset -r_\ell M_N^{\on} Y_\nu^{\on\, \dagger}= -\tfrac{1}{2} \big(r_\ell M_N^{\on} Y_\nu^{\on\, \dagger} + Y_\nu^{\on} M_N^{\on\, \dagger} r_\ell^\dagger \big) - \tfrac{1}{2} \big(r_\ell M_N^{\on} Y_\nu^{\on\, \dagger} - Y_\nu^{\on} M_N^{\on\, \dagger} r_\ell^\dagger \big),
	\end{equation}
having decomposed the counterterm into a Hermitian and an anti-Hermitian part for clarity. 

If we had consistently included the contributions from the redundant couplings to the field anomalous dimension~\eqref{eq:on-shell_gamma_formula} of the full on-shell formulation, we would have found 
	\begin{equation}
	\gamma^{(2)}_{1,\ell} =  Y_\nu^\on  \big( C_{H N}^{\on\, \dagger} M_N^\on - M^{\on \, \dagger}_N C_{ H N}^\on \big) Y_\nu^{\on\, \dagger}\\
	\end{equation}
instead of~\eqref{eq:ell_ADM_div}. The Hermitian contribution from the redundant coupling contribution to the wave-function renormalization~\eqref{eq:Zl_contribution} cancels the Hermitian~\eqref{eq:ell_ADM_div}, leaving only an anti-Hermitian piece. The divergence of the field anomalous dimension is now (a representation of) an element of the Lie algebra of the $ \U(3)_\ell $ flavor group associated with $ \ell $.
There is no change to the Yukawa coupling \bef.
We observe that the divergence of the \bef satisfies 
	\begin{equation} \label{eq:beta_Ynu_RG-fin}
	\beta^{(2),Y^\on_\nu}_{\on,1} = -\gamma^{(2)}_{1,\ell} Y^\on_\nu = -(\gamma^{(2)}_{1} \cdot Y^\on_\nu)
	\end{equation}
and conclude that the RG functions are indeed RG-finite when calculated consistently in the full on-shell formulation. Again, this formulation resolves the problems of the truncated on-shell RG functions.

One might wonder as to the nature of the observed divergence~\eqref{eq:beta_Ynu_div} of the Yukawa \bef in the truncated on-shell formulation. Should it be interpreted as being part of a divergent but RG-finite set of RG functions, or should we think of this as a spurious consequence of neglecting redundant couplings in the calculation? The latter interpretation seems very reasonable in light of~\eqref{eq:beta_Ynu_RG-fin}. 
In this view, however, the reader may find it puzzling why the RG functions prove to be divergent---albeit RG-finite---already at the two-loop order, whereas~\cite{Herren:2021yur} indicates that such divergences would typically enter only starting the three-loop order. It is not that the argument in~\cite{Herren:2021yur} is invalid because we are dealing with an EFT instead of a renormalizable theory; rather, the contribution from the redundant coupling to $ Z^{(1)}_{1,\ell} $~\eqref{eq:Zl_contribution} is generically non-Hermitian (contrary to the assumptions behind the three-loop argument in~\cite{Herren:2021yur}). Viewing the divergence as a spurious consequence of having neglected the contribution from the redundant coupling instead is similarly valid. The two perspectives are tied through the counterterm ambiguity described in Section~\ref{sec:RG_divergence}.

If we had worked in the truncated on-shell scheme, not including the effects of the redundant couplings, we would not see the $ r_\ell $-contribution to $ Z^{(1)}_{1,\ell} $. In the absence of this coupling, we might reasonably conclude that $ Z_{\ell} $ was chosen Hermitian and then expect that the RG functions come out finite at the two-loop order, contrary to observations. The apparent Hermitianity of $ Z_{\ell} $ is nothing but our ignorance of the redundant couplings. To actually force $ Z_{\ell} $ to be Hermitian, a divergent flavor rotation is needed to remove the anti-Hermitian part of the $ Z_\ell $ contribution~\eqref{eq:Zl_contribution}, but at the cost of introducing $ r_\ell $-dependence to the $ \delta Y_\nu $ counterterm. Only then can we apply the argument of~\cite{Herren:2021yur} that divergent RG functions are absent at the two-loop order. Now, however, the contribution from the redundant couplings to the \bef is not properly accounted for in the truncated on-shell formulation of the EFT. We would then interpret the divergence of the \bef as a spurious artifact from applying the incomplete RG formulas of the truncated formulation, i.e., neglecting the $ \beta_{\red,m}^\beta \partial_\beta $ term in~\eqref{eq:phys_beta-function_formula}.

\section{Geometry of the Coupling Space} \label{sec:geometry}
The various unphysical ambiguities plaguing the RG functions---even in the on-shell formulation, which is supposedly (more) physical---motivate us to explore the mathematical structure underlying the theory space and the RG functions. While we do not foresee immediate practical applications, it may, perhaps, provide a starting point for a deeper geometric exploration of the RG flow. If nothing else, it highlights the fact that when comparing different RG function calculations, one must take care only to compare the physical part lest one ends up chasing fictitious sources of errors.

\subsection{Revisiting flavor rotations and the coupling spaces}
It will prove useful to denote the (left) group action on the off-shell coupling space with the map $ L:\, G_F \times V_{\off} \to V_{\off} $ defined by $ L(a, \lambda) = L_a(\lambda) = a\cdot \lambda $ for $ a\in G_F $. The group action closes in $ V_\on $, so the restrictions $ L_a\big|_{V_\on} $ are automorphisms of the on-shell coupling space. We expect all functions in the $ V_\off $ and $ V_\on $ spaces to be $ G_F $-equivariant (casually referred to as covariance in physics parlance): at the most basic, the maps between $ V_\off $ and $ V_\on $ should ``commute'' with flavor rotations, meaning that 
	\begin{equation} \label{eq:on/off_map_equivariance}
	g_\Xi \circ L_a = L_a \circ g_\Xi, \qquad \lambda_\Xi \circ L_a = L_a \circ \lambda_\Xi,
	\qquad \forall a\in G_F.
	\end{equation}

We have repeatedly emphasized that physics is unchanged by flavor rotations. This implies a redundancy in our description of the theory even in the on-shell coupling space. The action of the flavor group on the coupling space entails a quotient space $ V_{\on} / G_F $ consisting of equivalence classes 
	\begin{equation}
	[g]_{G_F} = \big\{a\cdot g: a\in G_F \big\} \in  V_{\on} / G_F
	\end{equation} 
for $ g \in V_\on $.\footnote{The flavor action on $ V_\off $ also establishes a quotient space $ V_\off/G_F $; however, this space is still redundant under field-redefinitions, and we will not examine it closer here.} The invariance of physics under the flavor group implies that it should be possible to write all observables as functions on $ V_\on /G_F $. We are left to interpret $ V_\on /G_F $ as a \emph{physical} coupling space in the sense that there is no remnant redundancy in this space: each point describes unique physics.\footnote{The space $ V_\on /G_F $, on the other hand, is not itself unique but depends on the original choice $ V_\on \subseteq V_\off $.} The significance of the physical coupling space in the SM is discussed in Note~\ref{note:SM_phys_pars}.

\begin{note}[note:SM_phys_pars]{Physical parameters of the SM}
	For concreteness, the reader may think about the familiar case of the SM. In this case $ V_\on / G_F  $ can be parametrized (locally at least) by 18 physical parameters (three gauge couplings, nine diagonal Higgs couplings, four CKM parameters, and two parameters of the Higgs potential). By contrast, the space $ V_\on $ has an additional 41 physically redundant parameters in the three $ 3\times 3  $ Yukawa coupling matrices. In principle, one should also include the topological $ \theta $ terms; however, they do not transform linearly under the flavor group and one would have to extend our framework further.
\end{note}

The coupling spaces we have introduced in this article are related by
	\begin{equation}
	\vcenter{\hbox{\begin{tikzpicture} \pgfsetlinewidth{.6pt} 
	\node (a) at (0,.7) {$ V_\off $};
	\node (b) at (0,-.7) {$ V_\on \times V_\red $};
	\node (c) at (2,0) {$ V_\on $};
	\node (d) at (4,0) {$ V_\on/G_F $};
	\draw[->, transform canvas={shift={(.07,0)}}] (a.south) -- (b.north);
	\draw[->, transform canvas={shift={(-.07,0)}}] (b.north) -- (a.south);
	\draw[->] (a.south east) + (0,.2) --node[above,scale=.75,xshift= 2, yshift=-1] {$ g_\Xi $} ($(c.north west) + (0,-.15)$);
	\draw[->] (b.north east) + (0,0) --node[below,scale=.75,xshift= 2, yshift=2] {$ \psi $} ($(c.south west) + (0,.2)$);
	\draw[->] (c.east) --node[above,scale=.75] {$ \pi $} (d.west);
	\end{tikzpicture}}}.
	\end{equation}
With $ V_\off \cong V_\on\times V_\red $, we restrict our attention to just $ V_\off $ in this section as a familiar coupling space in which the vacuum functional can be renormalized and all RG functions determined. Parallel considerations apply to the full on-shell formulation in $ V_\on\times V_\red $. We, thus, focus on the sequence of coupling spaces
	\begin{equation}
	V_\off \xrightarrow{\;\; g_\Xi \;\; } V_\on \xrightarrow{\;\;\pi \;\;} V_\on / G_F,
	\end{equation}
where $ g_\Xi $ is the particular projection following from the choice $ \Xi(\eta, \lambda) $ of field redefinitions. The canonical map $ \pi:\, g \mapsto [g]_{G_F} $ projects to the physical space $ V_\on / G_F $. 
Clearly, $ \pi $ is invariant w.r.t. the group action:
	\begin{equation}
	\pi \circ L_a = \pi, \qquad \forall a\in G_F
	\end{equation}
following from the identification of $ G_F $-orbits with $ G_F $--equivalence class. Both $ g_\Xi $ and $ \pi $ are \emph{projections} in the sense that they admit right inverses. Whereas $ V_\off $ and $ V_{\on} $ are manifolds, $ V_\on / G_F $ needs only be an orbifold (which for the most part behaves as a manifold). We will examine its structure in Section~\ref{sec:geometry_of_physical_space}. 

From a geometric point of view, the \befs define vector fields on the coupling spaces, describing the flow of the couplings with changing renormalization scale. As we discussed in Section~\ref{sec:prelims},
the off-shell RG evolution of the theory is determined directly from the counterterms, needed for renormalizing the Green's functions. This allows for the calculation of an off-shell \bef, $ \beta_\off \in \mathfrak{X}(V_\off) $, the set of vector fields on $ V_\off $. It is well-known that \befs are $ G_F $-equivariant, which takes the form 
	\begin{equation}
	\beta_\off \circ L_a = L_{a\ast} \beta_\off, \qquad \forall a\in G_F.
	\end{equation}
Whereas we would not typically compute an on-shell \bef directly from counterterms, we described in Section~\ref{sec:beta_functions} how one might assign one by mapping between the on- and off-shell coupling space. Taking again $ \lambda_{\Xi}: V_\on \to V_\off $ to be a right-inverse of the projection $ g_\Xi $, the construction 
	\begin{equation} \label{eq:beta_on_def}
	\beta_\on = g_{\Xi\ast} (\beta_\off \circ \lambda_{\Xi}) \in \mathfrak{X}(V_\on)
	\end{equation}
defines a suitable on-shell \bef, where $ g_{\Xi\ast}: T(V_\off) \to T(V_\on) $ is the differential map (pushforward) between the tangent spaces. 

The off- and on-shell \befs are ambiguous in several ways. Firstly, $ \beta_\off $ is not uniquely defined due to the counterterm ambiguity in the renormalized theory. Other valid instances are of the form [cf.~\eqref{eq:RG_ct_ambiguity}]
	\begin{equation} \label{eq:beta_off_redef}
	\beta_\off^{\omega}=  \beta_\off - \omega^\sharp,
	\end{equation}
for a $ G_F $-equivariant function $ \omega: V_\off \to \mathfrak{g}_F $. Here $ \omega^\sharp: V_\off \to T(V_\off) $ denotes the fundamental vector field associated with $ \omega $ [written in terms of the Lie algebra action as $ \omega^\sharp(\lambda) = \omega(\lambda) \cdot \lambda $]. The same counterterm ambiguity also appears in the on-shell \befs, where~\eqref{eq:beta_on_ct_ambiguity} implies that 
	\begin{equation}
	\beta_\on^{\omega}=  \beta_\on - (\omega\circ \lambda_{\Xi})^\sharp,
	\end{equation}
where $ (\omega\circ \lambda_{\Xi})^\sharp : V_\on \to T(V_\on) $ denotes the fundamental vector field on the on-shell space. 

The \bef ambiguity from the counterterms lies in the vertical tangent space w.r.t.~$ \pi $. An integral curve generated by the action of a Lie algebra element lies in the $ G_F $ orbit and, so,\footnote{While $ V_\on/G_F $ is not generally a manifold, some submanifolds $ V\subseteq V_\on $ (e.g., the dense and connected principal stratum) are fiber bundles $ V \xrightarrow{\;\; \pi\;\;} V/G $. On such subspaces the kernel of the projection defines the vertical tangent spaces $ \ker(\pi_\ast|_V) =\mathrm{Ver}(V) $ and the fundamental vector field is $ \omega^\sharp: V \to \mathrm{Ver}(V) $.}
	\begin{equation}
	\omega^\sharp(\lambda_\Xi(g)) \in \ker(\pi_\ast|_g).
	\end{equation}
Hence, the ambiguity will vanish under a projection of $ \beta_\on $ to the physical coupling space. This is nothing but a formalization of the observation that arbitrary flavor rotations along the RG do not change physics~\cite{Fortin:2012hn,Jack:2013sha,Baume:2014rla,Herren:2021yur}.

A second source of arbitrariness in the definition of $ \beta_\on $ lies in the choice of field redefinitions informing the projection $ g_\Xi $. We could instead have used $ g_{\Xi_U} $ from the $ \Xi_U $ field redefinition~\eqref{eq:field_redef_ambiguity}. The difference can be parametrized by some function $ U: V_\off \to G_F $ that is equivariant in the sense that 
	\begin{equation} \label{eq:equivariant_GF_function}
	U\circ L_a = \mathrm{ad}_a \circ U,
	\end{equation}
where $ \mathrm{ad}_a: \, a' \mapsto a\, a' a^{\eminus 1} $ is the adjoint map of $ G_F $ defined by $ a \in G_F$. The shift in on-shell \befs associated with the new choice of projectors was derived in coordinate space in~\eqref{eq:beta_Xi_U_2}, and we identify the new \bef with  
	\begin{equation} \label{eq:beta_on_prime_U}
	\beta_{\on}^U \equiv g_{\Xi_U\ast} (\beta_\off \circ \lambda_{\Xi_U})= \beta_\on + \big(U^\ast \theta (\beta_\off \circ \lambda_\Xi) \big)^\sharp,
	\end{equation} 
where $ \theta \in \Omega^1(G_F) \otimes \mathfrak{g}_F $ is the Lie algebra--valued Maurer--Cartan one-form on $ G_F $ and $ U^\ast $ the pullback of $ U $. The contraction with $ \beta_\off \circ \lambda_\Xi $ gives a Lie algebra--valued function on $ V_\on $, which then induces a fundamental vector field on $ V_\on $ capturing the shift in $ \beta_\on $. The ambiguity due to the projection choice is a vertical vector field---an element of $ \ker(\pi_\ast) $---similarly to the one from the choice of counterterms and similarly unphysical.

\subsection{Limited projectability of the off-shell \bef}
We have seen several examples of how the on-shell \bef is not uniquely defined but, rather, depends on the specific choices for the off-shell counterterms and the projection $ g_{\Xi} $ (field redefinition $ \Xi $). This serves to reemphasize that $ V_\on $ is not a physical space. Only an RG flow in $ V_\on/G_F $ is unique and, in some loose sense, physical. Leaving aside for the moment that the physical space is not necessarily a manifold everywhere, we seek to obtain a physical \bef $ \beta_\mathrm{phys} \in \mathfrak{X}(V_\on/G_F) $. With $ \beta_\on $ $ G_F $-equivariant, it is necessarily \emph{projectable}\footnote{A vector field $ X $ on a manifold $ M $ is said to be projectable along the projection $ \varphi:M\to N $ if $ \varphi_\ast X_{p} = \varphi_\ast X_{p'} $ for all $ p,p'\in M$ such that $\varphi(p) = \varphi(p') $; that is, if it is possible to assign a unique value to the projected vector field for each point in $ N $. If $ X $ is projectable, $ \varphi_\ast X $ is well defined and equal to $ \varphi_\ast (X\circ f) $, for any right-inverse $ f $: $ \varphi\circ f = \mathrm{id}_N $.} to $ V_\on/G_F $, and we let 
	\begin{equation} 
	\beta_\mathrm{phys} = \pi_\ast \beta_\on.
	\end{equation}
Both the counterterm and projection-choice ambiguities are of the form $ \beta'_\on - \beta_\on \in \ker(\pi_\ast) $, as they only generate flow in the direction of a flavor rotation. 
Reassuringly, the ambiguities of $ \beta_\on $ are completely absent in this definition of $ \beta_\mathrm{phys} $, lending credence to the notion that this is indeed a physical \bef. 

The attentive reader might have caught on to one additional ambiguity in our construction of $ \beta_\on $: for physics to be consistently described across the off-shell theory space, it would seem that one could use the value of $ \beta_\off $ at any parameter point that projects down to the same on-shell coupling point. In light of the other ambiguities in $ \beta_\on $, we speculate that $ \beta_\off $ is not actually projectable to $ \beta_\on $ and, therefore, that $ g_{\Xi\ast} \beta_\off $ is ill defined. In this event, the definition~\eqref{eq:beta_on_def} depends on the particular choice of right inverse $ \lambda_{\Xi} $. In particular, we have in mind that different off-shell coupling points that project to the same on-shell point could result in $ \beta_\on $ values that differ in the direction of some flavor rotations. Physics is nevertheless uniquely defined as long as $ \beta_\mathrm{phys} $ is uniquely defined. Hence, we posit that $ \beta_\off $ is projectable all the way down to $ V_\on/G_F $ with $ \pi_{\ast}g_{\Xi\ast} $. In that event,
	\begin{equation}
	\beta_\mathrm{phys} = \pi_\ast g_{\Xi\ast} \beta_\off
	\end{equation} 
is well defined and unique: it will not depend on any choices made during the course of the calculations. Were $ \beta_\off $ not projectable to $ V_\on/G_F $, the consequences would be dire for our understanding of EFTs. Off-shell parameter points that are physically equivalent at one renormalization scale could then flow to different physics at other scales, a flagrant absurdity.

	\begin{figure}[t]
	\centering
	\includegraphics[width=.8\textwidth]{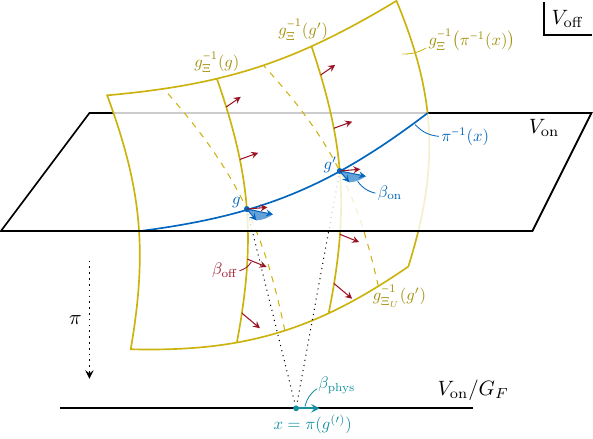}
	\caption{Sketch of the projection of the off-shell \bef vector field (red arrows) down to the physical subspace $ V_\on/G_F $ for the common choice $ g_{\Xi}|_{V_\on} = \mathrm{id}_{V_\on} $. The \bef vector field is not generally projectable onto $ V_\on $, meaning that different points along the  preimage $ g^{\eminus 1}_{\Xi}(g) $ project to different values in $ T_g(V_\on) $, illustrated with the blue cone. The projections of the vector field from anywhere on the yellow equivalence surface all map to a single vector in $ T_x(V_\on/G_F) $. }
	\label{fig:flow_projection}
	\end{figure}

Figure~\ref{fig:flow_projection} illustrates the \befs across the three coupling spaces, $ V_\off $, $ V_\on $, and $ V_\on/G_F $ with the typical choice $ g_{\Xi}|_{V_\on} = \mathrm{id}_{V_\on} $, so it behaves as an ordinary projection to a subspace. A particular physical point $ x\in V_\on/G_F $ has a preimage $ \pi^{\eminus 1}(x) $ of physically equivalent on-shell points, shown as a blue line in $ V_\on $. All points in the preimage $ g_{\Xi}^{\eminus1} (\pi^{\eminus 1}(x)) $, shown as a yellow surface in $ V_\off $, are yet again physically equivalent. Each point $ g $ in the on-shell equivalence class has a surface $ g_\Xi^{\eminus 1}(g) $ of equivalent off-shell points (illustrated as yellow lines). The ambiguity in choosing the projection $ g_{\Xi_U} $ is that of associating different equivalence classes (dashed yellow lines) to $ g $. The off-shell \bef is shown as red arrows on the equivalence surface, but we have not shown the non-uniqueness of $ \beta_\off $ so as not to clutter the figure. The conjectured non-projectability of $ \beta_\off $ to $ V_\on $ means that the values of $ \beta_\off $ on different points in $ g_\Xi^{\eminus 1}(g) $ project to different values for $ \beta_\on(g) $, illustrated by a blue cone of possible values. The ambiguities from choice of $ \beta_\off $ and projection $ g_\Xi $ are included in that cone too. Regardless of the ill-defined projection of $ \beta_\off $ to $ V_\on $, it is expected to be projectable down to $ \beta_\mathrm{phys} $ in $ V_\on/G_F $, yielding a unique value to the flow.

\subsection{Geometry of the physical coupling space} \label{sec:geometry_of_physical_space}
One caveat in the explanation above is that $ V_\on/G_F $ is not necessarily a manifold. This complicates the notion of establishing the vector field $ \beta_\mathrm{phys} $ in the physical coupling space. It also makes it impossible to interpret $ V_\on \xrightarrow{\;\pi \;} V_\on / G_F $ as a fiber bundle. A possible resolution is to \emph{stratify $ V_\on $ by orbit types} and focus on submanifolds for which the fiber bundle construction is possible. 

The flavor group $ G_F $---defined as the largest symmetry group of $ S_\mathrm{kin} $---does not generally act \emph{freely}\footnote{A group $ G $ is said to act \emph{freely} on a manifold $ M $ iff $ \forall p\in M, a\in G:\, a\cdot p = p \Leftrightarrow a = e $, with $ e\in G $ being the identity element.} on all of $ V_{\on} $ (or $ V_\off $). The implication is that not all orbits in $ V_\on $ by the group action are isomorphic to $ G_F $: some orbits are smaller. The would-be fibers from different base space points in $ V_\on/G_F $ are then not isomorphic, which prohibits the fiber bundle construction. An unmistakable example of this arises from all group elements acting trivially on the free theory, so the orbit is $ G_F\cdot 0 = \{0\} $---the free theory is maximally symmetric. More generally, the problem is the existence of points in coupling space that realize an enhanced symmetry. 

The \emph{stabilizer}\footnote{Also known as \emph{isotropy group} or \emph{symmetry group}.} $ G_g\subseteq G_F $ of a point $ g \in V_\on $ is defined by $ G_g = \big\{a \in G_F: a\cdot g = g \big\} $; it is the subgroup that acts trivially on $ g $, essentially the symmetry group at that coupling point. 
The stabilizers of the points along the orbit $ G_F\cdot g $ are conjugate to each other: we have
	\begin{equation}
	g' = a \cdot g \implies G_{g'} = a G_{g} a^{\eminus 1}
	\end{equation}  
for $ a \in G_F $ and $ g\in V_\on $. We write $ H \sim H' $ for conjugate subgroups $ H,H' \subseteq G_F $. Consequently, the orbit $ G_F \cdot g $ is characterized by a unique \emph{stabilizer conjugacy class} denoted $ (G_g) $. The orbit $ G_F \cdot g \cong G_F/G_g $~\cite[theorem~4.2.4]{Pflaum:2001}, meaning that the orbit can be identified with the coset of the stabilizer. Consequently, the different orbit types in $ V_\on $ are characterized by the stabilizer conjugacy classes. The subset of all \emph{fixed points} of subgroup $ H\subseteq G $ in $ V_\on $ is referred to as
	\begin{equation}
	V_{\on,H} = \big\{ g\in V_\on \, :\, G_g = H \big\}
	\end{equation} 
while the set of all points that share the stabilizer conjugacy class is 
	\begin{equation}
	V_{\on,(H)} = \big\{ g\in V_\on \, :\, G_g \sim H \big\}.
	\end{equation} 
If $ g \in V_{\on,(H)} $ for some subgroup $ H $ then the orbit $ G_F\cdot g \subseteq V_{\on,(H)} $.\footnote{We can also view every instance of a theory with an assumed global symmetry $ H $ as a particular restriction of a generic theory $ V_\on $ to the subspace $ V_{\on}^H = \big\{ g\in V_\on \, :\, G_g \supseteq H \big\} $.}

The idea now is to restrict our attention to a particular orbit type $ V_{\on,(H)} \subseteq V_\on $---as opposed to $ V_\on $ in its entirety---because these spaces form fiber bundles. It holds~\cite[theorem~IV.3.3]{Bredon:1972} that if $ V_{\on,(H)} $ is non-empty for some subgroup $ H\subseteq G_F $, then $ V_{\on,(H)} $ is a manifold and $ V_{\on,(H)} \xrightarrow{\;\pi\;} V_{\on,(H)} /G_F $ is a fiber bundle with a fiber isomorphic to $ G_F/H $. The structure group of the bundle is $ N_{G_F}\!(H)/H $, with $ N_{G_F}\!(H) $ being the \emph{normalizer}\footnote{The normalizer can be defined as $ N_{G_F}\!(H) = \big\{a \in G_F \, :\, aH a^{\eminus 1} = H \big\}$ and is the largest subgroup of $ G_F $ in which $ H $ is normal.} group of $ H $ in $ G_F $. These, then, are the structures needed to interpret the RG flow in a geometric context. 

An important theorem~\cite[theorems~4.3.2 and~IV.3.1, respectively]{Pflaum:2001,Bredon:1972}---sometimes known as the Principal Orbit Theorem---governs the $ G_F $-orbits in $ V_\on $. It states that there exists a unique conjugacy class $ (H^{\circ}) $ for which any other conjugacy class $ (H) $ emerging as a stabilizer of a $ g\in V_\on $ is larger, in the sense that $ H^{\circ} $ is conjugate to a subgroup of $ H $. The $ (H^{\circ}) $ orbit type is referred to as the \emph{principal orbit type} with $ V_{\on,(H^{\circ})} $ the \emph{principal orbit bundle}, and it follows that the dimension of a principal orbit is maximal among all the orbit types. It holds that $ V_{\on,(H^{\circ})} \subseteq V_{\on} $ is open and dense and, moreover, that $ V_{\on,(H^{\circ})}/G_F $ is connected. 
The intuition for our purposes is that a ``typical'' or ``generic'' point in $ V_\on $ will be part of a maximal orbit, which is isomorphic to $ G_F/H^{\circ} $. This is where one would expect to find a physics model if no additional symmetries have been imposed on the theory. Clearly, there are also \emph{singular orbits} (e.g., $ G_F\cdot 0 $) that have lower dimensionality than the principal ones.\footnote{There may also be \emph{exceptional orbits} which are of the same dimension but are covering maps of the principal ones.} To provide context, Note~\ref{note:SM_normalizer} exemplifies these constructions for the familiar case of the SM. 

\begin{note}[note:SM_normalizer]{Flavor and symmetry groups of the SM}
	The flavor group of the SM is $ G_F= \U(3)^5 $. This is generally broken by the Yukawa couplings, which are the only parameters transforming non-trivially under $ G_F $. The couplings realized in nature---with non-degenerate, nonzero masses, and a non-trivial CKM matrix---violates $ G_F $ maximally; there is no choice of Yukawa matrices that would break $ G_F $ to a smaller symmetry group. We conclude that the version of the SM realized in nature belongs to the principal orbit space with the stabilizer $ H^{\circ}= \U(1)_{B+L} \times U(1)_{B-L} \times \U(1)_{L_\mu - L_e} \times \U(1)_{L_\tau-L_\mu} $, of which the first factor is anomalous. 
	
	The $ U(1)_{B\pm L} $ factors are central subgroups of $ G_F $, so all quark flavor transformations are part of the normalizer $ N\!(H^\circ) $. To determine the normalizer in the lepton sector, note that $ H^\circ a\cdot g = a H^\circ \cdot g = a \cdot g $ for $ a \in N\!(H^\circ) $ and $ g \in V_{\on,H^\circ} $, so the action of any element of the normalizer on a fixed point of $ H^\circ $ results in another fixed point. One can then verify that 
		\begin{equation*}
		N\!(H^\circ) = \U(3)_q \times \U(3)_u \times \U(3)_d \times \!\! \left( \prod_{f\in\{e,\mu,\tau\}} \!\! \U(1)_{\ell_f} \times \U(1)_f \! \right) \!\! \rtimes S_3,
		\end{equation*}
	where $ S_3 $ is the permutation group of the three lepton families. Meanwhile, the centralizer of $ H^\circ $ is 
		\begin{equation*}
		C\!(H^\circ) = \U(3)_q \times \U(3)_u \times \U(3)_d \times \!\! \left( \prod_{f\in\{e,\mu,\tau\}} \!\! \U(1)_{\ell_f} \times \U(1)_f \! \right).
		\end{equation*}
\end{note}

It turns out that it is sufficient to focus on a single orbit type, say $ (H) $, when analyzing the RG trajectory of a theory. The physics case for this statement is that the RG cannot enhance or decrease the symmetry of a theory along its flow. We can show that this is the case by considering some coupling point $ g\in V_\on $. The RG flow generates an integral curve $ \gamma: I \to V_\on $, for some open interval $ I \subseteq \mathbb{R} $, such that $ \dot{\gamma}(t) = \beta_\on(\gamma(t)) $ and $ \gamma(0) = g $. Since $ \beta_\on $ is $ G_F $-equivariant, we have that  
	\begin{equation}
	\dfrac{\dd}{\dd t}(a\cdot \gamma)= L_{a\ast} \dot{\gamma} = L_{a\ast} \beta_\on(\gamma) = \beta_\on(a\cdot \gamma)
	\end{equation}
for any $ a\in G_F $. Thus, $ a\cdot \gamma $ is an integral curve generated by $ \beta_{\on} $ as well. In the case that $ a\in G_g $, we see that $ a\cdot \gamma(0) = \gamma(0) $, so $ a\cdot \gamma $ is a solution to the same ordinary differential equation as $ \gamma $ and with the same initial value. By uniqueness of the solution, we conclude that $ a \cdot \gamma $ and $ \gamma $ are identical integral curves. In particular, this indicates that 
	\begin{equation}
	a \cdot \gamma(t) = \gamma(t), \qquad \forall a \in G_g.
	\end{equation}
For the stabilizer this implies that $ G_g \subseteq G_{\gamma(t)} $ for all points along the flow. Nor can we flow to a point of increased symmetry along the flow: Assume for contradiction that $ \exists t \in \mathbb{R}: G_{\gamma(t)} \supsetneq G_g $, then $ g = \gamma'(\eminus t) $ for the integral curve $ \gamma' $ of $ \beta_\on $ with initial value $ \gamma'(0) = \gamma(t) $, which implies $ G_{\gamma(t)} \subseteq G_g $. 

The above argument for the flow generated by $ \beta_\on $ indicates that an integral curve $ \gamma $ with initial value in $ V_{\on,H} $ for some stabilizer $ H \subseteq G_F $ satisfies $ \gamma(I) \subseteq V_{\on,H} $, such that the flow is entirely contained in the set of $ H $ fixed points. This shows that it is sufficient to consider a suitable stabilizer conjugacy class $ V_{\on,(H)} $ when examining the flow generated by $ \beta_\on $, which in particular facilitates the use of the $ V_{\on,(H)} \xrightarrow{\;\pi\;} V_{\on,(H)} /G_F $ bundle. One should keep in mind that the integral curve might still asymptote to a point of enhanced symmetry not in $ V_{\on,H} $.\footnote{Instead the asymptotic point would lie in $ V_\on^H = \{g\in V_\on \, :\, G_g \supseteq H \} \supseteq V_{\on,H}$.} This could happen, for instance, when a flow asymptotes to the Gaussian fixed point (which has maximal symmetry and minimal orbit). 

Finally, let us consider what the \bef ambiguities look like on the restricted coupling spaces. 
It is especially relevant that a $ G_F $-equivariant function $U: V_\on \to G_F $ cannot take arbitrary values: If we consider a point $ g\in V_\on $,  equivariance~\eqref{eq:equivariant_GF_function} implies that
	\begin{equation}
	a U(g) = U(a\cdot g) a= U(g) a, \qquad \forall a\in G_g.
	\end{equation}
It follows that $ U(g) \in C_{G_F}\!(G_g) $, the \emph{centralizer}\footnote{The centralizer $ C_{G_F}\!(H) \subseteq N_{G_F}\!(H) $ for a subgroup $ H \subseteq G_F $ and is defined by $ C_{G_F}(H) = \{b\in G_F\, : \, ba=ab\; \forall a \in H\} $.} of $ G_g $ in $ G_F $. The image of any equivariant $ G_F $-valued function $ U $ restricted to a set of fixed points satisfies
	\begin{equation}
	U(V_{\on,H}) \subseteq C_{G_F}\!(H).
	\end{equation}
A similar argument can be extended to the level of $ G_F $-equivariant Lie algebra--valued functions (such as appears in the \bef ambiguities), say $ \omega:V_\on \to \mathfrak{g}_F $. Letting $ \mathfrak{h}\subseteq \mathfrak{g}_F $ be the subalgebra associated with $ H $, we have 
	\begin{equation} \label{eq:gF_equivariant_function}
	\omega(V_{\on,H}) \subseteq \mathfrak{z}_{\mathfrak{g}_F}(\mathfrak{h}), \qquad \mathfrak{z}_{\mathfrak{g}_F}(\mathfrak{h}) = \big\{\alpha \in \mathfrak{g}_F\,:\, [\alpha, \beta] = 0\; \forall \beta \in \mathfrak{h} \big\}
	\end{equation} 
Here $ \mathfrak{z}_{\mathfrak{g}_F}(\mathfrak{h}) $ is the centralizer of $ \mathfrak{h} $ in $ \mathfrak{g}_F $. All \bef ambiguities we have identified are generated by fundamental vector fields associated with such Lie algebra--valued functions and are, therefore, constrained by the result~\eqref{eq:gF_equivariant_function}.

\subsection{In search of a unique lift of the physical \bef}
So far our approach has been to go from unphysical \befs to increasingly more physical ones, but what about going the other way? It may not be terribly convenient to work directly in the physical quotient space $ V_\on/G_F $, so we might wish to lift $ \beta_\mathrm{phys} $ up to $ V_\on $, essentially to find a suitable and, ideally, unique representative element. The previous discussion emphasizes that $ V_\on \to V_\on/G_F $ is not a fiber bundle.\footnote{Barring some trivial cases like $ G_F = 1 $.} Instead, we need to commit to a particular orbit type, say $ H\subseteq G_F $, with the fiber bundle $ V_{\on,(H)} \xrightarrow{\;\pi\;} V_{\on,(H)} /G_F $. Within such a space, the question of lifting $ \beta_\mathrm{phys} $ is well posed. Additionally, we can be certain that the flow generated by the RG will be contained within this bundle. 
Any of the ambiguous \befs $ \beta_\on \in \mathfrak{X}(V_{\on,(H)}) $ project down to $ \beta_\mathrm{phys} $ on the base space and can be seen as particular choices of lifts from $ \beta_\mathrm{phys} $. The challenge is then to fix a unique such choice.

Defining a \emph{horizontal lift} of a vector field in a fiber bundle means prescribing a separation of the tangent bundle of the total space $ T(V_{\on,(H)}) = \Ver(V_{\on,(H)}) \oplus \mathrm{Hor}(V_{\on,(H)})$, where the vertical subspace is identified by $ \Ver(V_{\on,(H)}) = \ker(\pi_{\ast}|_{V_{\on,(H)}}) $, the vectors that project to zero on the base. An Ehresmann connection on the bundle is a choice for the horizontal subspace $ \mathrm{Hor}(V_{\on,(H)}) $ and is equivalent to choosing a connection $ \Phi \in  \Omega^{1}(V_{\on,(H)}) \otimes T(V_{\on,(H)}) $---a tangent space--valued one-form---acting as a projector onto the vertical subspace. It presents as a somewhat daunting task to uniquely define a connection one-form. Indeed, it does not appear to be generally possible to construct a $ G_F $-equivariant connection one-form. We may be rescued by the fact that the \bef is not fully arbitrary within $ \Ver(V_{\on,(H)}) $. 

The vertical subspace consists of fundamental vector fields $ \alpha^{\sharp} $ generated by Lie algebra--valued functions $ \alpha: V_{\on,(H)} \to \mathfrak{g}_F $ (or, rather, for every vertical vector field, there is such a Lie algebra--valued function). By restricting our consideration to $ G_F $-equivariant functions, we strongly constrain the functions $ \alpha $. Let w.l.o.g. $ g \in V_{\on,H} \subseteq V_{\on,(H)} $. It follows from~\eqref{eq:gF_equivariant_function} that an equivariant $ \alpha $ imposes $ \alpha(g) \in \mathfrak{z}_{\mathfrak{g}_F}(\mathfrak{h}) $ with $ \mathfrak{h} $ the Lie algebra of $ H $. Furthermore, $ \mathfrak{h} $ is the kernel of the vector field map $ \sharp: \mathfrak{g}_F \to \Ver_g(V_{\on,(H)}) $, so it is sufficient to consider the values
	\begin{equation}
	\alpha(g) \in \mathfrak{z}_{\mathfrak{g}_F}(\mathfrak{h}) / \mathfrak{h}.
	\end{equation}
The ambiguity in the on-shell \befs is similarly generated by equivariant Lie algebra--valued functions and must be of the same form. Hence, the condition
	\begin{equation} \label{eq:connection_criteria}
	\Phi|_g \circ \sharp \in \mathrm{Aut}\big( \mathfrak{z}_{\mathfrak{g}_F}(\mathfrak{h}) / \mathfrak{h} \big), \qquad \forall g \in V_{\on,H}
	\end{equation}
is a sufficient requirement for an equivariant connection: $ \Phi $ then allows for the construction of an invertible function on the space of equivariant fundamental vector fields. The requirement 
	\begin{equation} \label{eq:horizontal_lift_condition}
	\Phi(\beta_\on) = 0
	\end{equation}
will then uniquely fix a horizontal, $ G_F $-equivariant lift of $ \beta_\mathrm{phys} $.

We take inspiration from the special case of renormalizable field theories, where $ V_\on = V_\off $. Here a candidate connection one-form presents itself. There is a unique and finite \bef, known as the flavor-improved \bef, $ B_\off \in \mathfrak{X}(V_\off)$~\cite{Fortin:2012hn,Jack:2013sha,Baume:2014rla,Herren:2021yur}, which does not suffer from ambiguities due to the choice of renormalization constants unlike the other \befs. 
Furthermore there is a gauge (choice of counterterms) for which $ \beta_\off = B_\off $, so $ \pi_\ast B_\off = \beta_\mathrm{phys} $ for renormalizable theories. Thus, $ B_\off $ is an obvious candidate for a unique lift of $ \beta_\mathrm{phys} $. There exists another unique RG function $ P_\off \in \Omega^1(V_\off) \otimes \mathfrak{g}_F $---a Lie algebra--valued one-form---associated with the RG evolution of background gauge fields associated with the flavor group~\cite{Fortin:2012hn,Jack:2013sha,Baume:2014rla}.\footnote{$ P $ (our $ P_\off $) is referred to as $ \tilde{\rho} $ in~\cite{Jack:2013sha}.} Also $ P_\off $ is invariant under a change of counterterms, so it is uniquely defined. It was found that the requirement 
	\begin{equation} \label{eq:P_B_contraction}
	P_\off(B_\off) = P_{\off,i} B_\off^{i} = 0
	\end{equation}
is enforced by a consistency condition arising from the requirement that Weyl transformations (local scale transformations) form an Abelian group. 

We find condition~\eqref{eq:P_B_contraction} tantalizing in its similarity to~\eqref{eq:horizontal_lift_condition}. We speculate that $ P_\off $ may satisfy the criteria~\eqref{eq:connection_criteria}. If this is borne out, $ P_\off $ constitutes an unambiguous $ G_F $-equivariant connection one-form on the $ V_{\off,(H)} \xrightarrow{\;\pi\;} V_{\off,(H)} /G_F $ bundle. Furthermore, of all possible $ \beta_\off $, the flavor-improved \bef $ B_{\off} $ would be the horizontal lift of $ \beta_\mathrm{phys} $ in renormalizable theories. With $ B_{\off} $ being finite in the $ \epsilon $-expansion~\cite{Herren:2021yur}, this manifests the finiteness of $ \beta_\mathrm{phys} $ as well. These are all very desirable properties of a horizontal lift.

While we find the above structure compelling for renormalizable theories, further calculations are required to test, if not prove, this hypothesis. It also leaves the question of whether one can assert a unique connection one-form in $ V_{\on,(H)} \xrightarrow{\;\pi\;} V_{\on,(H)} /G_F $ in EFTs. The flavor-improved RG functions $ P_{\off} $ and $ B_{\off} $ have, to our knowledge, been used and examined only in renormalizable theories so far. While they should generalize to off-shell or full on-shell formulations of EFTs, it is generally not possible to pullback $ P_\off $ to the on-shell space. Perhaps, one could hope that the partial factorization of on-shell and redundant couplings in the full on-shell formulation means that $ P_{\on \times \red} $ is a \emph{basic} one-form, ensuring that it is itself a pullback of a form $ P_\on $. Additional research is needed to clarify the picture.

\section{Conclusion} \label{sec:conclusions}
We have examined the origin of curious divergences observed in EFT RG functions calculated recently~\cite{Jenkins:2023bls,Manohar:2024xbh,Zhang:2025ywe,Naterop:2025cwg} with the truncated on-shell formulation---also known as an ``on-shell basis''---at the two-loop order. The bare Lagrangian of this framework does not renormalize the Green's functions of the theory, and the resulting RG functions are generally ill-defined; they do not describe the flow of the Green's functions under the RG. A resolution to recover a well-behaved RG flow for the vacuum functional is to extend the on-shell formulation with NMSTs, with which the vacuum functional can be fully renormalized. In this formulation, the on-shell RG functions are rendered RG-finite.  

The off-shell and full on-shell formulations of an EFT are entirely equivalent, with an invertible map between the couplings of the two. While the full on-shell formulation does not provide a reduction in the number of couplings as does the truncated on-shell formulation, it exhibits a separation of on-shell and redundant couplings absent in the off-shell formulation. The \befs of the on-shell couplings are shown to be independent of the redundant couplings with the exception of unphysical contributions that generate flavor rotations along the RG flow. 

Practical calculations of the RG flow of on-shell couplings can be performed in the truncated on-shell formulation for computational convenience; however, the \befs may exhibit divergences that generate unphysical flavor rotations during the RG flow. In contrast to the full formulations~\cite{Herren:2021yur} (on and off shell), these divergences occur starting at the two-loop order. In the truncated on-shell framework, `t Hooft consistency conditions for the on-shell \befs are valid only up to violations proportional to an element of $ \mathfrak{g}_F $ acting on the couplings. It is always possible to perform a divergent flavor rotation to remove any divergence from the on-shell \befs in both the truncated and the full on-shell frameworks.

We have observed that unphysical flavor rotations prevent the interpretation of the on-shell coupling space as being physical; flavor transformations lay bare extra redundancies in this space. The consequence at the level of on-shell \befs is that they are ambiguous in the direction aligned with orbits generated by the flavor group. In particular they are sensitive to choices of counterterms, the choice of projection from the off-shell space, and the choice of right inverse (embedding) of the on-shell theory. We have taken some initial steps towards establishing a \bef in a physical coupling space (the coset of the on-shell w.r.t. the flavor group) and analyzed how \befs in the various coupling spaces are related. This provides an initial geometric framing of the RG in coupling space. It remains to be seen to what extent this construction is practical. Another interesting topic for further studies is what changes with inclusion of couplings that transform non-linearly under the flavor group (topological terms).

\subsection*{Acknowledgments}
I am grateful to Lukas Born, Javier Fuentes-Martín, Julie Pagès, Peter Stoffer, and Javier Virto for interesting discussions inspiring this work and to Lukas Born, Joe Davighi, and Julie Pagès for giving feedback on early drafts. This work was funded by the Swiss National Science Foundation (SNSF) through the Ambizione grant ``Matching and Running: Improved Precision in the Hunt for New Physics,'' project number 209042.

%\renewcommand{\thesection}{\Alph{section}}
%\appendix

\sectionlike{References}
\vspace{-10pt}
\bibliography{References} 
\end{document}

%% file: PreRamble.tex
\usepackage[english]{babel}
\usepackage{microtype,xspace}
\usepackage[utf8]{inputenc}	% Allows for writing special charachters in the tex-file 

\usepackage{amsfonts,amsmath,amssymb,bm,mathrsfs,mathtools,dsfont} 	% Standard mathematics 
\allowdisplaybreaks 	%Allows pagebreak during multiline math enviroments
\usepackage[table]{xcolor}
\usepackage{hyperref}
\usepackage{slashed,cancel}
\usepackage{multicol}

\usepackage{siunitx}
\usepackage{geometry}
\usepackage[sort&compress,numbers,merge]{natbib}
\usepackage{chapterbib}
\addto\captionsenglish{\renewcommand*{\bibname}{References}}
\makeatletter
\renewcommand{\@memb@bchap}{ %\section*{\bibname} 
\bibmark \prebibhook
}
\makeatother

\usepackage{graphicx}
\graphicspath{{./Figures/}}
\usepackage{caption,subcaption}
\usepackage{multirow}

\usepackage{tabularx}
\newcolumntype{Y}{>{\centering\arraybackslash}X}

\usepackage[shortlabels]{enumitem}
\setlist{itemsep=.1em,topsep=.5em}
\SetEnumerateShortLabel{i}{\textit{\roman*}}

\definecolor{red}{rgb}{0.6,.0706,.1373}
\definecolor{blue}{rgb}{0,0.396,0.741}
\definecolor{green}{rgb}{0.25,0.6,0.2}
\definecolor{teal}{rgb}{0.11,0.6,0.6}
\definecolor{orange}{rgb}{.8, .4806, 0.173}
\definecolor{yellow}{rgb}{.8, .7, 0.05}
\colorlet{blueref}{blue!80!black}
\colorlet{bluelink}{blue!90!black}
\hypersetup{
	colorlinks, 
	bookmarksopen, 
	bookmarksnumbered,
	citecolor=blueref, 		%color of links to bibliography
	linkcolor=bluelink,	%color of internal links
	urlcolor=bluelink,			%color of external links 
	pdftitle={\thetitle},    %   - title (PDF meta)
	pdfauthor={\theauthor},    	%   - author (PDF meta)
}

\usepackage{tikz}
\usetikzlibrary{positioning,calc,arrows.meta}

\addto\captionsenglish{}

\renewcommand{\contentsname}{Contents}
\renewcommand{\printtoctitle}[1]{}

\settocdepth{subsection}
\newcommand{\toc}{ {
	\hypersetup{linkcolor = black} %Locally changes the linkcolor for the toc
	\vspace*{-.06\textheight}	
	\tableofcontents*
	\thispagestyle{standardstyle} 
} }

\makeatletter
\newcommand*\ifthispageodd{%
  \checkoddpage
  \ifoddpage
    \expandafter\@firstoftwo
  \else
    \expandafter\@secondoftwo
  \fi
}
\makeatother

\usepackage[noindentafter]{titlesec} %Removes the indentation after new section
\counterwithout{section}{chapter} % Include if sections should be taken at the top level
\counterwithout{figure}{chapter} %Removes chapter number on figures
\counterwithout{table}{chapter} %Removes chapter number on tables
\numberwithin{equation}{section} % Sets numbering at the section level
\setsecnumdepth{subsubsection}

\DeclareMathVersion{sans}
\SetSymbolFont{operators}{sans}{OT1}{cmbr}{m}{n}
\SetSymbolFont{letters}{sans}{OML}{cmbrm}{m}{it}
\SetSymbolFont{symbols}{sans}{OMS}{cmbrs}{m}{n}
\SetMathAlphabet{\mathit}{sans}{OT1}{cmbr}{m}{sl}
\SetMathAlphabet{\mathbf}{sans}{OT1}{cmbr}{bx}{n}
\SetMathAlphabet{\mathtt}{sans}{OT1}{cmtl}{m}{n}
\SetSymbolFont{largesymbols}{sans}{OMX}{iwona}{m}{n}

\DeclareMathVersion{boldsans}
\SetSymbolFont{operators}{boldsans}{OT1}{cmbr}{b}{n}
\SetSymbolFont{letters}{boldsans}{OML}{cmbrm}{b}{it}
\SetSymbolFont{symbols}{boldsans}{OMS}{cmbrs}{b}{n}
\SetMathAlphabet{\mathit}{boldsans}{OT1}{cmbr}{b}{sl}
\SetMathAlphabet{\mathbf}{boldsans}{OT1}{cmbr}{bx}{n}
\SetMathAlphabet{\mathtt}{boldsans}{OT1}{cmtl}{b}{n}
\SetSymbolFont{largesymbols}{boldsans}{OMX}{iwona}{bx}{n}

\titleformat{\section}{\needspace{3\baselineskip} \centering \Large \bfseries \sffamily \mathversion{boldsans} \color{blue!80!black} }{\thesection}{15pt}{}{}
\titlespacing{\section}{0pt}{15pt}{5pt}
\titleformat{\subsection}{\needspace{2\baselineskip} \large \sffamily \mathversion{sans} \color{blue!70!black} }{\thesubsection}{10pt}{}{}
\titlespacing{\subsection}{0pt}{12pt}{5pt}
\titleformat{\subsubsection}{\normalsize \sffamily \itshape \mathversion{sans} \color{blue!70!black} }{\thesubsubsection}{10pt}{}{}
\titlespacing{\subsubsection}{0pt}{10pt}{0pt}

\newcommand{\sectionlike}[1]{\phantomsection \addcontentsline{toc}{section}{#1} \setcounter{subsection}{0} \sectionmark{#1}
		\begin{center}
		\needspace{8\baselineskip}
		\Large \bfseries \sffamily \mathversion{boldsans} \color{blue!80!black} #1  
		\end{center}
	\vspace{-5pt} 
}

\ifnum\fullheadfoot=1
	\makepagestyle{standardstyle}
	\makeoddhead{standardstyle}{\sffamily \mathversion{subsectionmath} \rightmark}{}{\sffamily \bfseries{\thepage}}
	\makeevenhead{standardstyle}{\sffamily \bfseries{\thepage}}{}{\sffamily \mathversion{subsectionmath} \leftmark}
	\makeheadrule{standardstyle}{\textwidth}{\normalrulethickness}
	\makepsmarks{standardstyle}{
		\nouppercaseheads
		\createmark{section}{both}{shownumber}{\ }{\hspace{1.2em}  }
		\createmark{subsection}{right}{shownumber}{}{\hspace{1.2em} }
	}
	
	\copypagestyle{appstyle}{standardstyle}
	\makeevenhead{appstyle}{\sffamily \bfseries{\thepage}}{}{ \sffamily \mathversion{subsectionmath} \leftmark}
	\makepsmarks{appstyle}{
		\nouppercaseheads
		\createmark{section}{both}{nonumber}{\ }{\ \ }
		\createmark{subsection}{right}{shownumber}{}{\ \ }
	}
\else %Plain style
	\makepagestyle{standardstyle}
	\makeoddfoot{standardstyle}{}{\sffamily -- {\bfseries\thepage} --}{} 
	\makeevenfoot{standardstyle}{}{\sffamily -- {\bfseries\thepage} --}{}
	\copypagestyle{appstyle}{standardstyle}
\fi

\createplainmark{toc}{both}{\contentsname}
\createplainmark{bib}{both}{\bibname}

\makeatletter
\let\MyIntOrig\int
\def\MyIntSpace{\hspace{-.35em}} %% Configure as needed.
\def\int{\MyInt}
\def\MyInt{\MyIntOrig\MyIntSkipMaybe}
\def\MyIntSkipMaybe{
	\@ifnextchar_{\MyIntSkipScript}{%
		\@ifnextchar^{\MyIntSkipScript}{%
			\@ifnextchar\limits{\MyIntSkipTok}{%
				\@ifnextchar\nolimits{\MyIntSkipTok}{%
					\MyIntSpace}}}}%
}
\def\MyIntSkipScript#1#2{#1{#2}\MyIntSkipMaybe}
\def\MyIntSkipTok#1{#1\MyIntSkipMaybe}

\newcommand{\pushright}[1]{\ifmeasuring@#1\else\omit\hfill$\displaystyle#1$\fi\ignorespaces}
\makeatother

\usepackage[most,many,breakable]{tcolorbox}
\tcbset{shield externalize} %Makes the boxes play nice with tikzexternal

\tcbuselibrary{skins}
\makeatletter
\newtcolorbox[auto counter%,number within=section
	]{note}[2][Note]{enhanced,
	breakable,
	before upper={\parindent15pt\noindent}, %allow for regular paragraph indentation
	pad before break= 2mm,
	pad after break= 2mm,
	float=t,
	colback=white,
	colframe=blue!80!black!70,
	attach boxed title to top left={yshift*=-\tcboxedtitleheight},
	fonttitle=\bfseries \sffamily,
	label= #1,
	title={Note~\thetcbcounter:
	\mathversion{boldsans} #2},
	boxed title size=title,
	boxed title style={%
	        right=-1pt,
	        sharp corners,
	        rounded corners=northwest,
	        colback=tcbcolframe,
	        boxrule=0pt,
	    },
	underlay boxed title={%
	        \path[fill=tcbcolframe] (title.south west)--(title.south east)
	        to[out=0, in=180] ([xshift=5mm]title.east)--
	        (title.center-|frame.east)
	        [rounded corners=\kvtcb@arc] |-
	        (frame.north) -| cycle;
	    },
}
\makeatother

%% file: References.bib
@article{Anselmi:2012aq,
    author = "Anselmi, Damiano",
    title = "{A General Field-Covariant Formulation Of Quantum Field Theory}",
    eprint = "1205.3279",
    archivePrefix = "arXiv",
    primaryClass = "hep-th",
    reportNumber = "IFUP-TH-2012-08",
    doi = "10.1140/epjc/s10052-013-2338-5",
    journal = "Eur. Phys. J. C",
    volume = "73",
    number = "3",
    pages = "2338",
    year = "2013"
}

@article{Arzt:1993gz,
    author = "Arzt, Christopher",
    title = "{Reduced effective Lagrangians}",
    eprint = "hep-ph/9304230",
    archivePrefix = "arXiv",
    reportNumber = "UM-TH-92-28",
    doi = "10.1016/0370-2693(94)01419-D",
    journal = "Phys. Lett. B",
    volume = "342",
    pages = "189--195",
    year = "1995"
}

@article{Baume:2014rla,
    author = "Baume, Florent and Keren-Zur, Boaz and Rattazzi, Riccardo and Vitale, Lorenzo",
    title = "{The local Callan-Symanzik equation: structure and applications}",
    eprint = "1401.5983",
    archivePrefix = "arXiv",
    primaryClass = "hep-th",
    doi = "10.1007/JHEP08(2014)152",
    journal = "JHEP",
    volume = "08",
    pages = "152",
    year = "2014"
}

@article{Bednyakov:2014pia,
    author = "Bednyakov, A. V. and Pikelner, A. F. and Velizhanin, V. N.",
    title = "{Three-loop SM beta-functions for matrix Yukawa couplings}",
    eprint = "1406.7171",
    archivePrefix = "arXiv",
    primaryClass = "hep-ph",
    reportNumber = "HU-MATHEMATIK-2014-17, HU-EP-14-27",
    doi = "10.1016/j.physletb.2014.08.049",
    journal = "Phys. Lett. B",
    volume = "737",
    pages = "129--134",
    year = "2014"
}

@article{Bento:2023owf,
    author = "Bento, Miguel P. and Silva, Joao P. and Trautner, Andreas",
    title = "{The basis invariant flavor puzzle}",
    eprint = "2308.00019",
    archivePrefix = "arXiv",
    primaryClass = "hep-ph",
    doi = "10.1007/JHEP01(2024)024",
    journal = "JHEP",
    volume = "01",
    pages = "024",
    year = "2024"
}

@article{Buchmuller:1985jz,
    author = "Buchmuller, W. and Wyler, D.",
    title = "{Effective Lagrangian Analysis of New Interactions and Flavor Conservation}",
    reportNumber = "CERN-TH-4254/85",
    doi = "10.1016/0550-3213(86)90262-2",
    journal = "Nucl. Phys. B",
    volume = "268",
    pages = "621--653",
    year = "1986"
}

@article{Callan:1970yg,
    author = "Callan, Jr., Curtis G.",
    title = "{Broken scale invariance in scalar field theory}",
    doi = "10.1103/PhysRevD.2.1541",
    journal = "Phys. Rev. D",
    volume = "2",
    pages = "1541--1547",
    year = "1970"
}

@article{Caswell:1974cj,
    author = "Caswell, William E. and Wilczek, Frank",
    title = "{On the Gauge Dependence of Renormalization Group Parameters}",
    reportNumber = "Print-74-0568 (PRINCETON)",
    doi = "10.1016/0370-2693(74)90437-7",
    journal = "Phys. Lett. B",
    volume = "49",
    pages = "291--292",
    year = "1974"
}

@article{Chetyrkin:2017ppe,
    author = "Chetyrkin, K. G.",
    title = "{Combinatorics of $\mathbf{R}$-, $\mathbf{R^{-1}}$-, and $\mathbf{R^*}$-operations and asymptotic expansions of feynman integrals in the limit of large momenta and masses}",
    eprint = "1701.08627",
    archivePrefix = "arXiv",
    primaryClass = "hep-th",
    reportNumber = "PI-PH-PTH-13-91",
    month = "1",
    year = "2017"
}

@article{Criado:2018sdb,
    author = "Criado, J. C. and P{\'e}rez-Victoria, M.",
    title = "{Field redefinitions in effective theories at higher orders}",
    eprint = "1811.09413",
    archivePrefix = "arXiv",
    primaryClass = "hep-ph",
    doi = "10.1007/JHEP03(2019)038",
    journal = "JHEP",
    volume = "03",
    pages = "038",
    year = "2019"
}

@article{Deans:1978wn,
    author = "Deans, W. S. and Dixon, John A.",
    title = "{Theory of Gauge Invariant Operators: Their Renormalization and S Matrix Elements}",
    doi = "10.1103/PhysRevD.18.1113",
    journal = "Phys. Rev. D",
    volume = "18",
    pages = "1113--1126",
    year = "1978"
}

@article{Einhorn:2001kj,
    author = "Einhorn, Martin B. and Wudka, Jose",
    title = "{Effective beta functions for effective field theory}",
    eprint = "hep-ph/0105035",
    archivePrefix = "arXiv",
    reportNumber = "MCTP-01-20, UCRHEP-T310",
    doi = "10.1088/1126-6708/2001/08/025",
    journal = "JHEP",
    volume = "08",
    pages = "025",
    year = "2001"
}

@article{Einhorn:2013kja,
    author = "Einhorn, Martin B. and Wudka, Jose",
    title = "{The Bases of Effective Field Theories}",
    eprint = "1307.0478",
    archivePrefix = "arXiv",
    primaryClass = "hep-ph",
    reportNumber = "UCRHEP-T529, NSF-ITP-13-115, MCTP-13-18",
    doi = "10.1016/j.nuclphysb.2013.08.023",
    journal = "Nucl. Phys. B",
    volume = "876",
    pages = "556--574",
    year = "2013"
}

@article{Fortin:2012hn,
    author = "Fortin, Jean-Francois and Grinstein, Benjamin and Stergiou, Andreas",
    title = "{Limit Cycles and Conformal Invariance}",
    eprint = "1208.3674",
    archivePrefix = "arXiv",
    primaryClass = "hep-th",
    reportNumber = "UCSD-PTH-12-10, CERN-PH-TH-2012-297, SU-ITP-12-38",
    doi = "10.1007/JHEP01(2013)184",
    journal = "JHEP",
    volume = "01",
    pages = "184",
    year = "2013"
}

@article{Fuentes-Martin:2022jrf,
    author = {Fuentes-Mart{\'\i}n, Javier and K{\"o}nig, Matthias and Pag{\`e}s, Julie and Thomsen, Anders Eller and Wilsch, Felix},
    title = "{A proof of concept for matchete: an automated tool for matching effective theories}",
    eprint = "2212.04510",
    archivePrefix = "arXiv",
    primaryClass = "hep-ph",
    reportNumber = "MITP-22-105, TUM-HEP-1443/22, ZU-TH-58/22",
    doi = "10.1140/epjc/s10052-023-11726-1",
    journal = "Eur. Phys. J. C",
    volume = "83",
    number = "7",
    pages = "662",
    year = "2023"
}

@article{Georgi:1991ch,
    author = "Georgi, Howard",
    title = "{On-shell effective field theory}",
    reportNumber = "HUTP-91-A014",
    doi = "10.1016/0550-3213(91)90244-R",
    journal = "Nucl. Phys. B",
    volume = "361",
    pages = "339--350",
    year = "1991"
}

@article{Giudice:2007fh,
    author = "Giudice, G. F. and Grojean, C. and Pomarol, A. and Rattazzi, R.",
    title = "{The Strongly-Interacting Light Higgs}",
    eprint = "hep-ph/0703164",
    archivePrefix = "arXiv",
    reportNumber = "CERN-PH-TH-2007-47",
    doi = "10.1088/1126-6708/2007/06/045",
    journal = "JHEP",
    volume = "06",
    pages = "045",
    year = "2007"
}

@article{Henriksson:2025vyi,
    author = "Henriksson, Johan and Kousvos, Stefanos R. and Roosmale Nepveu, Jasper",
    title = "{EFT meets CFT: Multiloop renormalization of higher-dimensional operators in general $\boldsymbol{\phi^4}$ theories}",
    eprint = "2511.16740",
    archivePrefix = "arXiv",
    primaryClass = "hep-th",
    reportNumber = "CERN-TH-2025-258",
    month = "11",
    year = "2025"
}

@article{Herren:2017uxn,
    author = "Herren, Florian and Mihaila, Luminita and Steinhauser, Matthias",
    title = "{Gauge and Yukawa coupling beta functions of two-Higgs-doublet models to three-loop order}",
    eprint = "1712.06614",
    archivePrefix = "arXiv",
    primaryClass = "hep-ph",
    reportNumber = "TTP17-046",
    doi = "10.1103/PhysRevD.97.015016",
    journal = "Phys. Rev. D",
    volume = "97",
    number = "1",
    pages = "015016",
    year = "2018",
    note = "[Erratum: Phys.Rev.D 101, 079903 (2020)]"
}

@article{Herren:2021yur,
    author = "Herren, Florian and Thomsen, Anders Eller",
    title = "{On ambiguities and divergences in perturbative renormalization group functions}",
    eprint = "2104.07037",
    archivePrefix = "arXiv",
    primaryClass = "hep-th",
    reportNumber = "FERMILAB-PUB-21-196-T",
    doi = "10.1007/JHEP06(2021)116",
    journal = "JHEP",
    volume = "06",
    pages = "116",
    year = "2021"
}

@article{tHooft:1973mfk,
    author = "'t Hooft, Gerard",
    title = "{Dimensional regularization and the renormalization group}",
    doi = "10.1016/0550-3213(73)90376-3",
    journal = "Nucl. Phys. B",
    volume = "61",
    pages = "455--468",
    year = "1973"
}

@article{tHooft:1973wag,
    author = "'t Hooft, Gerard and Veltman, M. J. G.",
    title = "{DIAGRAMMAR}",
    reportNumber = "CERN-73-09",
    doi = "10.1007/978-1-4684-2826-1_5",
    journal = "NATO Sci. Ser. B",
    volume = "4",
    pages = "177--322",
    year = "1974"
}

@article{Jack:1990eb,
    author = "Jack, I. and Osborn, H.",
    title = "{Analogs for the $c$ Theorem for Four-dimensional Renormalizable Field Theories}",
    reportNumber = "DAMTP-90-02",
    doi = "10.1016/0550-3213(90)90584-Z",
    journal = "Nucl. Phys. B",
    volume = "343",
    pages = "647--688",
    year = "1990"
}

@article{Jack:2013sha,
    author = "Jack, I. and Osborn, H.",
    title = "{Constraints on RG Flow for Four Dimensional Quantum Field Theories}",
    eprint = "1312.0428",
    archivePrefix = "arXiv",
    primaryClass = "hep-th",
    reportNumber = "DAMTP-13-53",
    doi = "10.1016/j.nuclphysb.2014.03.018",
    journal = "Nucl. Phys. B",
    volume = "883",
    pages = "425--500",
    year = "2014"
}

@article{Jenkins:2023bls,
    author = "Jenkins, Elizabeth E. and Manohar, Aneesh V. and Naterop, Luca and Pag{\`e}s, Julie",
    title = "{Two loop renormalization of scalar theories using a geometric approach}",
    eprint = "2310.19883",
    archivePrefix = "arXiv",
    primaryClass = "hep-ph",
    reportNumber = "ZU-TH 69/23, PSI-PR-23-39",
    doi = "10.1007/JHEP02(2024)131",
    journal = "JHEP",
    volume = "02",
    pages = "131",
    year = "2024"
}

@article{Keren-Zur:2014sva,
    author = "Keren-Zur, Boaz",
    title = "{The local RG equation and chiral anomalies}",
    eprint = "1406.0869",
    archivePrefix = "arXiv",
    primaryClass = "hep-th",
    doi = "10.1007/JHEP09(2014)011",
    journal = "JHEP",
    volume = "09",
    pages = "011",
    year = "2014"
}

@article{Kluberg-Stern:1975ebk,
    author = "Kluberg-Stern, H. and Zuber, J. B.",
    title = "{Renormalization of Nonabelian Gauge Theories in a Background Field Gauge. 2. Gauge Invariant Operators}",
    reportNumber = "SACLAY-DPh-T/75/28",
    doi = "10.1103/PhysRevD.12.3159",
    journal = "Phys. Rev. D",
    volume = "12",
    pages = "3159--3180",
    year = "1975"
}

@article{Manohar:2024xbh,
    author = "Manohar, Aneesh V. and Pag{\`e}s, Julie and Roosmale Nepveu, Jasper",
    title = "{Field redefinitions and infinite field anomalous dimensions}",
    eprint = "2402.08715",
    archivePrefix = "arXiv",
    primaryClass = "hep-ph",
    reportNumber = "DESY-24-020, HU-EP-24/04-RTG",
    doi = "10.1007/JHEP05(2024)018",
    journal = "JHEP",
    volume = "05",
    pages = "018",
    year = "2024"
}

@article{Naterop:2024cfx,
    author = "Naterop, Luca and Stoffer, Peter",
    title = "{Renormalization-group equations of the LEFT at two loops: dimension-five effects}",
    eprint = "2412.13251",
    archivePrefix = "arXiv",
    primaryClass = "hep-ph",
    reportNumber = "PSI-PR-24-30, ZU-TH 66/24",
    doi = "10.1007/JHEP06(2025)007",
    journal = "JHEP",
    volume = "06",
    pages = "007",
    year = "2025"
}

@article{Naterop:2025cwg,
    author = "Naterop, Luca and Stoffer, Peter",
    title = "{Renormalization-group equations of the LEFT at two loops: dimension-six operators}",
    eprint = "2507.08926",
    archivePrefix = "arXiv",
    primaryClass = "hep-ph",
    reportNumber = "ZU-TH 48/25",
    month = "7",
    year = "2025"
}

@article{Symanzik:1970rt,
    author = "Symanzik, K.",
    title = "{Small distance behavior in field theory and power counting}",
    reportNumber = "DESY-70-20",
    doi = "10.1007/BF01649434",
    journal = "Commun. Math. Phys.",
    volume = "18",
    pages = "227--246",
    year = "1970"
}

@article{Zhang:2025ywe,
    author = "Zhang, Di",
    title = "{Two-loop renormalization group equations in the {\ensuremath{\nu}}SMEFT}",
    eprint = "2504.00792",
    archivePrefix = "arXiv",
    primaryClass = "hep-ph",
    doi = "10.1007/JHEP06(2025)106",
    journal = "JHEP",
    volume = "06",
    pages = "106",
    year = "2025"
}

@book{Collins:1984xc,
    author = "Collins, John C.",
    title = "{Renormalization : An Introduction to Renormalization, the Renormalization Group and the Operator-Product Expansion}",
    doi = "10.1017/9781009401807",
    isbn = "978-0-521-31177-9, 978-0-511-86739-2, 978-1-009-40180-7, 978-1-009-40176-0, 978-1-009-40179-1",
    publisher = "Cambridge University Press",
    address = "Cambridge",
    series = "Cambridge Monographs on Mathematical Physics",
    volume = "26",
    year = "1984"
}

@book{Pflaum:2001,
	author = {Pflaum, Markus J.},
	year = {2001},
	title = {Analytic and Geometric Study of Stratified Spaces},
	isbn = {978-3-540-42626-4},
	publisher = {Springer Berlin, Heidelberg},
	doi = "10.1007/3-540-45436-5"
}

@book{Bredon:1972,
	author = {Bredon, Glen},
	year = {1972},
	title = {Introduction to compact transformation groups},
	isbn = {9780080873596},
	publisher = {Academic Press},
}
